\documentclass[twocolumn]{aastex63}
\usepackage{multirow}
\usepackage[figuresright]{rotating}
\usepackage{subfigure}
\usepackage{epic,eepic}
\usepackage{graphicx}
\usepackage{longtable}
\usepackage{float}
\usepackage{lineno}
\usepackage{pifont}
\usepackage{gensymb}
\usepackage{graphicx}
\usepackage{threeparttable}
\usepackage{textcomp, gensymb}
\usepackage{amsmath}
\usepackage{color}
\usepackage{hyperref}
\newcommand{\uat}[2]{\href{http://astrothesaurus.org/uat/#2}{#1 (#2)}}

\shorttitle{\rm NUV emission evolution}
\shortauthors{Li et al.}

\begin{document}


\title{\rm Evolution of Stellar Activity and Habitable zone: I. Ultraviolet Emission of Dwarfs in Open Clusters and Field Stars}

\correspondingauthor{Song Wang}
\email{songw@bao.ac.cn}

\author{Xue Li}
\affiliation{Key Laboratory of Optical Astronomy, National Astronomical Observatories, Chinese Academy of Sciences, Beijing 100101, China}
\affiliation{School of Astronomy and Space Sciences, University of Chinese Academy of Sciences, Beijing 100049, China}

\author{Song Wang}
\affiliation{Key Laboratory of Optical Astronomy, National Astronomical Observatories, Chinese Academy of Sciences, Beijing 100101, China}
\affiliation{Institute for Frontiers in Astronomy and Astrophysics, Beijing Normal University, Beijing 102206, China}

\author{Jun Ma}
\affiliation{Key Laboratory of Optical Astronomy, National Astronomical Observatories, Chinese Academy of Sciences, Beijing 100101, China}
\affiliation{School of Astronomy and Space Sciences, University of Chinese Academy of Sciences, Beijing 100049, China}

\author{Henggeng Han}
\affiliation{Key Laboratory of Optical Astronomy, National Astronomical Observatories, Chinese Academy of Sciences, Beijing 100101, China}

\author{Yang Huang}
\affiliation{Key Laboratory of Optical Astronomy, National Astronomical Observatories, Chinese Academy of Sciences, Beijing 100101, China}
\affiliation{School of Astronomy and Space Sciences, University of Chinese Academy of Sciences, Beijing 100049, China}

\author{Jifeng Liu}
\affiliation{Key Laboratory of Optical Astronomy, National Astronomical Observatories, Chinese Academy of Sciences, Beijing 100101, China}
\affiliation{School of Astronomy and Space Sciences, University of Chinese Academy of Sciences, Beijing 100049, China}
\affiliation{Institute for Frontiers in Astronomy and Astrophysics, Beijing Normal University, Beijing 102206, China}
\affiliation{New Cornerstone Science Laboratory, National Astronomical Observatories, Chinese Academy of Sciences, Beijing 100012, People's Republic of China}

\let\cleardoublepage\clearpage
\begin{abstract}

Near-ultraviolet (NUV) radiation from dwarf stars plays a critical role in shaping the habitability of planetary systems, yet its long-term evolution across different spectral types remains poorly investigated. 
Based on GALEX NUV observations, we study the evolution of stellar NUV emission for a sample of 386,500 A- to M-type dwarfs spanning ages from 3 ~Myr to 10~Gyr, drawn from both open clusters and the field.
The normalized NUV emission ($f_{\rm NUV}/f_{\rm J}$) is used to trace the evolutionary trends. 
Our results reveal distinct evolutionary pathways after considering the distance completeness: A and early-F dwarfs show a weak decline in NUV emission during the main-sequence phase; late-F to G dwarfs exhibit a clear decrease, consistent with continuous spin-down driven by magnetic braking; late-K and M-dwarfs undergo a rapid decline in NUV emission when they evolve from young stellar objects to main-sequence stars.
Furthermore, we construct the evolutionary tracks of stellar ultraviolet habitable zone (UHZ). By comparing stellar circumstellar habitable zone (CHZ) and UHZ, we find that G- and K-type stars offer the most stable overlap between thermal and UV habitability over long-term evolution.

\end{abstract}


\section{INTRODUCTION}
\label{intro.sec}

The evolution of stellar activity is a central question in stellar astrophysics, with crucial implications for stellar age-dating, magnetic field evolution, and the habitability of exoplanets. This evolution can be investigated using various stellar activity tracers, such as Ca II H\&K, $\rm H\alpha$, X-ray, and ultraviolet (UV) emissions.

For decades, both the Ca II H\&K lines and X-ray emission have been widely used to study chromospheric and coronal activity, respectively. Ca II H\&K lines, for example, have line core emission strengths that serve as reliable proxies for stellar magnetic activity. Pioneering studies \citep[e.g.,][]{1972ApJ...171..565S, 1981ApJ...247..210B, 1991ApJ...375..722S, 2004A&A...426.1021P} have shown that Ca II H\&K activity index keeps declining with age, reflecting the weakening of stellar magnetic fields over time. 
However, \cite{2013A&A...551L...8P} noted that this activity index appears to plateau for stars older than a few gigayears, proposing an ``L-shaped" diagram of chromospheric activity versus age. In contrast, recent studies \citep[e.g.,][]{2008ApJ...687.1264M, 2016A&A...594L...3L, 2018A&A...619A..73L, 2024ApJS..271...19Y} suggest that the relationship between Ca II H\&K activity and age continues to decline, which contrasts with the results of \cite{2013A&A...551L...8P}. 
To resolve this discrepancy and clarify the long-term evolution of chromospheric activity, further studies based on well-characterized stellar samples spanning a broad range of ages are essential.

NUV emission plays a critical role in determining stellar habitability. The relationship between NUV emission and stellar age has been explored in several studies, particularly focusing on cooler stars like K and M dwarfs. For instance, the studies based on the GALEX and MUSCLES data suggest that NUV emission decreases with stellar age in K dwarfs \citep[e.g.,][]{2019ApJ...872...17R, 2022ApJ...929..169R, 2023ApJ...951...44R} and M dwarfs \citep[e.g.,][]{2013MNRAS.431.2063S, 2014AJ....148...64S, 2021ApJ...907...91L}. However, the relationship between NUV emission and age for F and G dwarfs remains poorly explored, limiting our understanding of the habitable zone evolution and the long-term habitability of planetary systems around these stars.

In this work, we investigate the near-ultraviolet (NUV) emission evolution for A to M dwarfs in open clusters based on Gaia DR2/DR3 and field stars based on the Large Sky Area Multi-Object Fiber Spectroscopic Telescope (LAMOST) DR10 data, using the photometric data from the Galaxy Evolution Explorer (GALEX) survey. In Section~\ref{samples.sec}, we present the collection and selection of samples from both open clusters and field stars. In Section \ref{uv_emission.sec}, we discuss the relation between NUV emission and stellar age in different spectral types, including A to M dwarfs. In Section \ref{uhz.sec}, we discuss the evolution of habitable zone with age. Finally, we summary our work in Section \ref{summary.sec}.

\section{Sample}
\label{samples.sec}

The samples we used in this study including the open clusters and filed stars. Here we describe the construction and selection of the two samples.  

\subsection{Open Cluster Sample}
\label{oc_sample.sec}

Member stars of open clusters share similar properties, such as age, chemical composition, extinction, and distance. This homogeneity enables precise age determination with the isochrone fitting method, thereby facilitating the investigation of how NUV radiation evolves with stellar age. In this study, we used open cluster samples compiled from multiple studies:

1) \cite{2020A&A...640A...1C} provided reliable parameters of 1,867 clusters identified using Gaia DR2 data by \cite{2018A&A...615A..49C}. The clusters in this catalog can be traced to approximately 4 kpc.

2) \cite{2024A&A...686A..42H} identified open clusters using advanced classification methods based on the classification criteria from \cite{2023A&A...673A.114H}, which identified 7,167 clusters using Gaia DR3 data, and provided a more stringent subset of 4,105 highly reliable clusters. Here we choose clusters with a color-magnitude diagram (CMD) classification percentile above 5 and a cluster significance based on astrometric signal-to-noise ratio (SNR) test greater than 0.5 to ensure robust detections. 

3) \cite{2020AJ....160..279K} provided 8,292 open clusters within 3 kpc based on Gaia DR2 data, some of which exhibit elongated, string-like structures. To ensure the reliability of our sample, we used on the labels provided in the paper to exclude such morphologies, removing approximately 4,000 clusters in which member stars could not be clearly distinguished from the surrounding field stars.

4) \cite{2022A&A...661A.118C} compiled open cluster data from their previous works \citep{2018A&A...618A..59C, 2019A&A...627A..35C, 2020A&A...635A..45C}, identifying a total of 1,274 open clusters using Gaia DR2 and eDR3 data, with most of them being located beyond 1 kpc from the Sun. 

5) \cite{2023ApJS..264....8H} discovered totally 2,541 star clusters using Gaia eDR3 data, with 1,656 new star cluster candidates beyond 1.2 kpc of the solar system.

6) \cite{2022A&A...660A...4H} detected 1,930 previously known open clusters, 82 known globular clusters, and 704 new stellar clusters as potential open clusters located at Galactic latitudes of $|b| \leq 20^\circ$, using Gaia eDR3 data.

For the studies by \cite{2020A&A...640A...1C} and \cite{2020AJ....160..279K}, which provided Gaia DR2 IDs, we cross-matched these Gaia DR2 IDs with the latest Gaia DR3 data using the table    ``dr3.dr2\_neighborhood'' from the Gaia Archive\footnote{\url{https://gea.esac.esa.int/archive/}}, obtaining corresponding Gaia DR3 IDs. Additionally, we retrieved the J2000 coordinates for each source using the CASJob tool available at NADC\footnote{\url{http://nadc.china-vo.org/casjobs/}}.

Since an open cluster may appear in multiple catalogs, the identified member stars and cluster parameters can vary between them. Figure \ref{hr_4_clusters_compare.fig} (in Appendix \ref{comOC.sec}) displays the CMD of several well-studied star clusters from \cite{2020A&A...640A...1C}, \cite{2024A&A...686A..42H} and \cite{2020AJ....160..279K}. The member stars from these studies show noticeable discrepancies. The clusters from \cite{2024A&A...686A..42H} include more members, especially cool stars. However, this may introduce contamination from field stars. In contrast, the cluster members identified by \cite{2020A&A...640A...1C} exhibit a much clearer and tighter main sequence, indicating higher reliability.
We also compared the cluster parameters provided by the different datasets, as shown in Figure \ref{oc_pars_compare.fig} (in Appendix \ref{comOC.sec}). The ages and distances reported by \cite{2020A&A...640A...1C} and \cite{2024A&A...686A..42H} are generally consistent, but there is a difference of approximately 0.4–0.5 magnitudes in extinction, which was also reported by \cite{2023A&A...673A.114H}. On the other hand, the ages derived by \cite{2020AJ....160..279K} are different from the other two datasets: for young clusters, \cite{2020AJ....160..279K} tend to report older ages, while for older clusters, they tend to report younger ages. The distance determined by \cite{2020AJ....160..279K} aligns well with the other datasets.
Therefore, after a comparison of the CMD and cluster parameters across multiple datasets, we prioritized the cluster parameters and member stars as follows:
\cite{2020A&A...640A...1C}, \cite{2024A&A...686A..42H}, \cite{2020AJ....160..279K}, \cite{2022A&A...661A.118C}, \cite{2023ApJS..264....8H}, and \cite{2022A&A...660A...4H}. For example, for the cluster Pleiades, we will only use the members and parameters from \cite{2020A&A...640A...1C}. We compiled a final catalog containing 900,300 member stars from open clusters.

\subsection{Field Star Sample}
\label{field_sample.sec}

Estimating stellar ages for field stars remains a long-standing challenge in stellar astrophysics, primarily due to the lack of direct and universally applicable age indicators. Traditional methods such as $isochrone$ fitting rely on precise determinations of stellar parameters, and are most effective for stars in rapid evolutionary phases, making them less suitable for unevolved dwarfs. Gyrochronology uses stellar rotation periods to infer ages, but it is limited to relatively young stars (age $\lesssim 4$~Gyr) \citep{2016Natur.529..181V}. Asteroseismology can yield precise ages but is only applicable to solar-like stars. 

In this study, the stellar ages for the field star sample are derived from a machine learning model trained on LAMOST spectra \citep{2025ApJS..280...13W}. The model is trained using wide binaries with known ages, field stars with ages inferred from chemical abundance-based methods, and cluster stars, and applied to LAMOST DR10 to yield a large catalog of age estimates for over 4 million dwarf stars. 
The age uncertainty of this method decreases from 30–50\% for younger stars ($\lesssim$2 Gyr) to $\simeq$15$\%$ for stars around 10~Gyr. In addition, F- and G-type stars generally show smaller uncertainties compared to K- and M-type stars. As an example, the age uncertainty is about $10\%-25\%$ for K-type stars with the SNR above 50 \citep{2025ApJS..280...13W}.
We selected stars with spectra that have high SNR, i.e., SNR$_i >10$ and SNR$_g >$10. To ensure reliable age estimates, we selected only stars with ages between 1 and 10~Gyr.


\begin{figure*}
    \centering
    \subfigure[]{
    \includegraphics[width=0.46\textwidth]{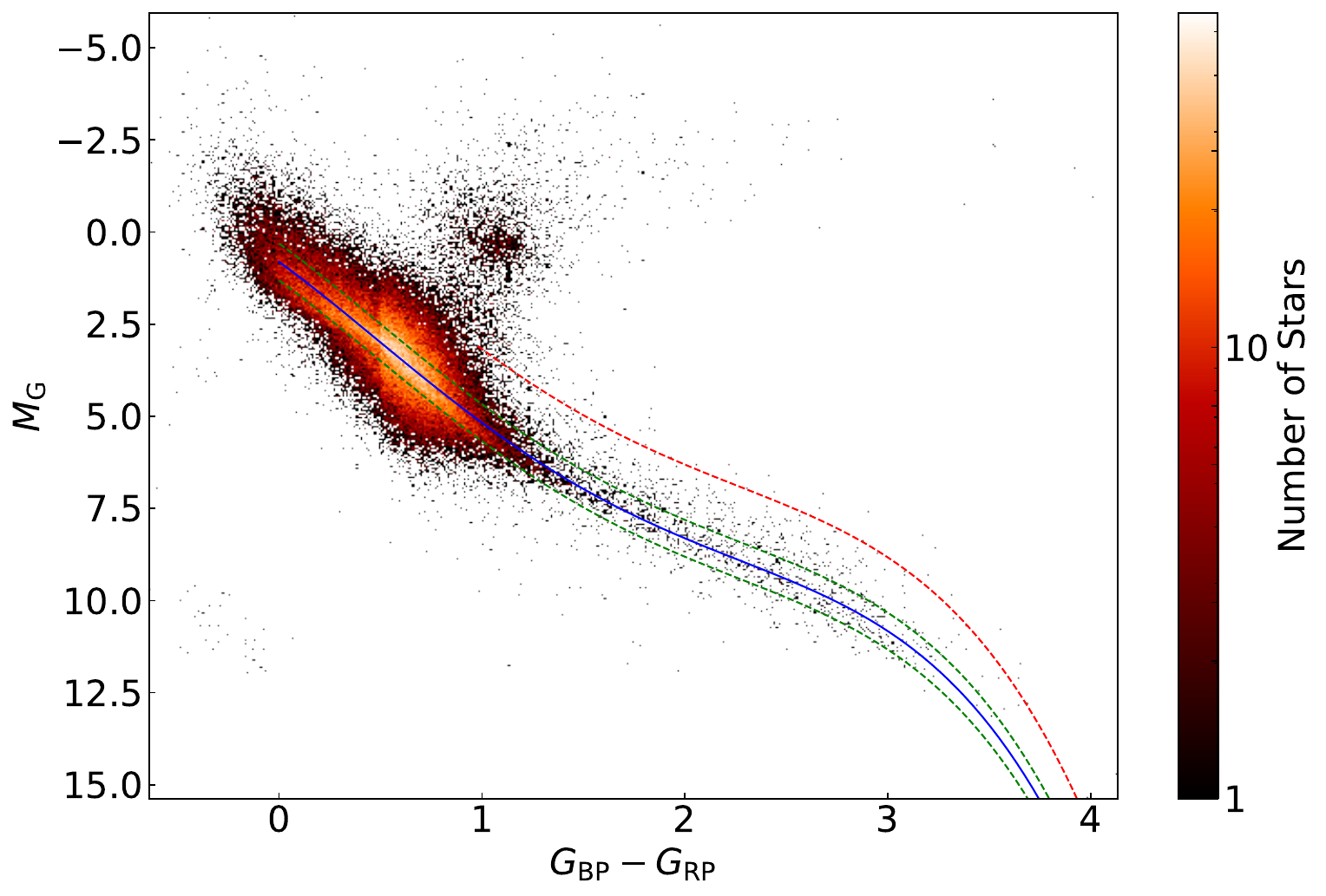}
    \label{hr_gaia_oc.fig}}
    \subfigure[]{
    \includegraphics[width=0.46\textwidth]{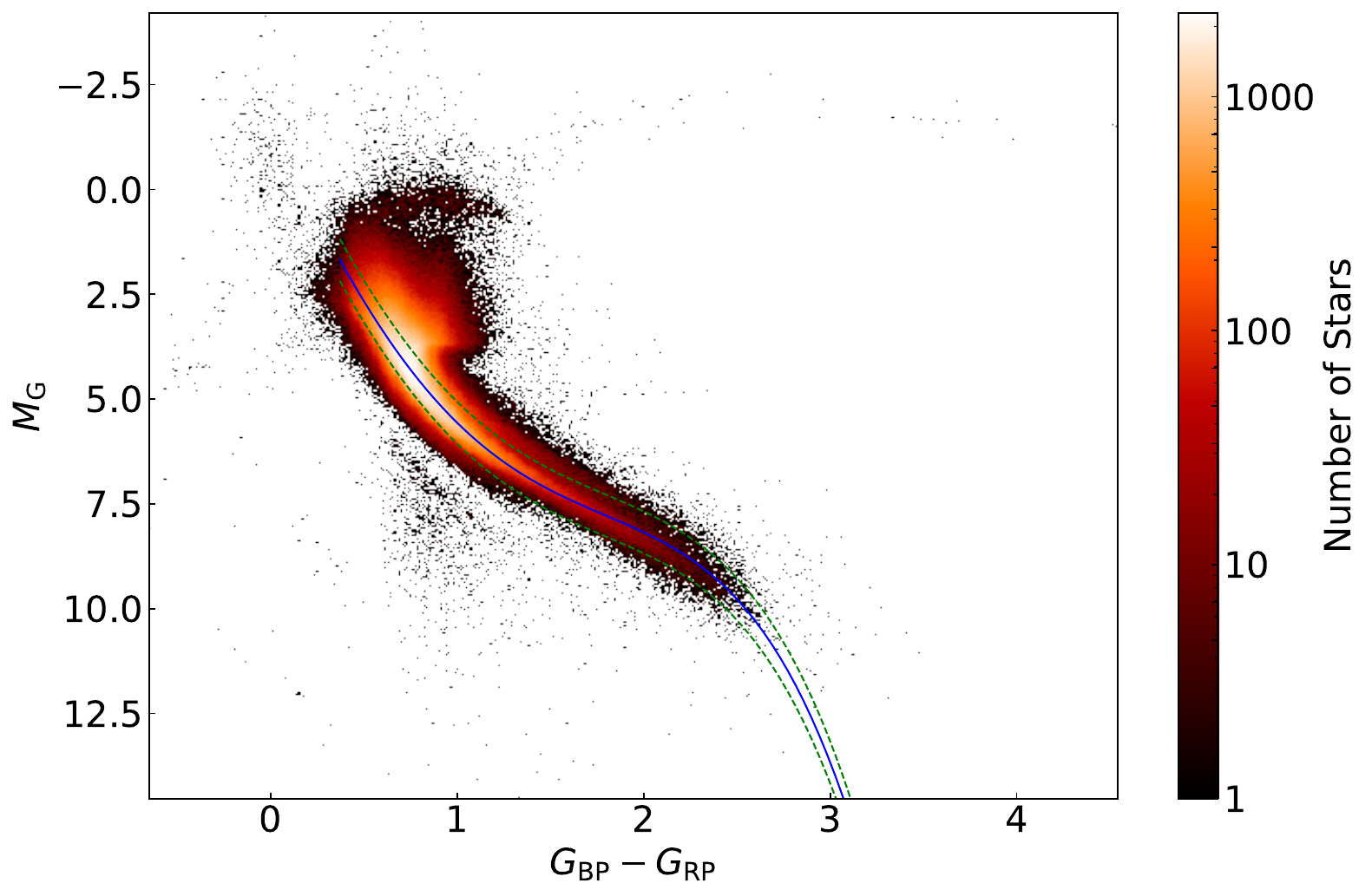}
    \label{hr_gaia_field.fig}}
    \subfigure[]{
    \includegraphics[width=0.46\textwidth]{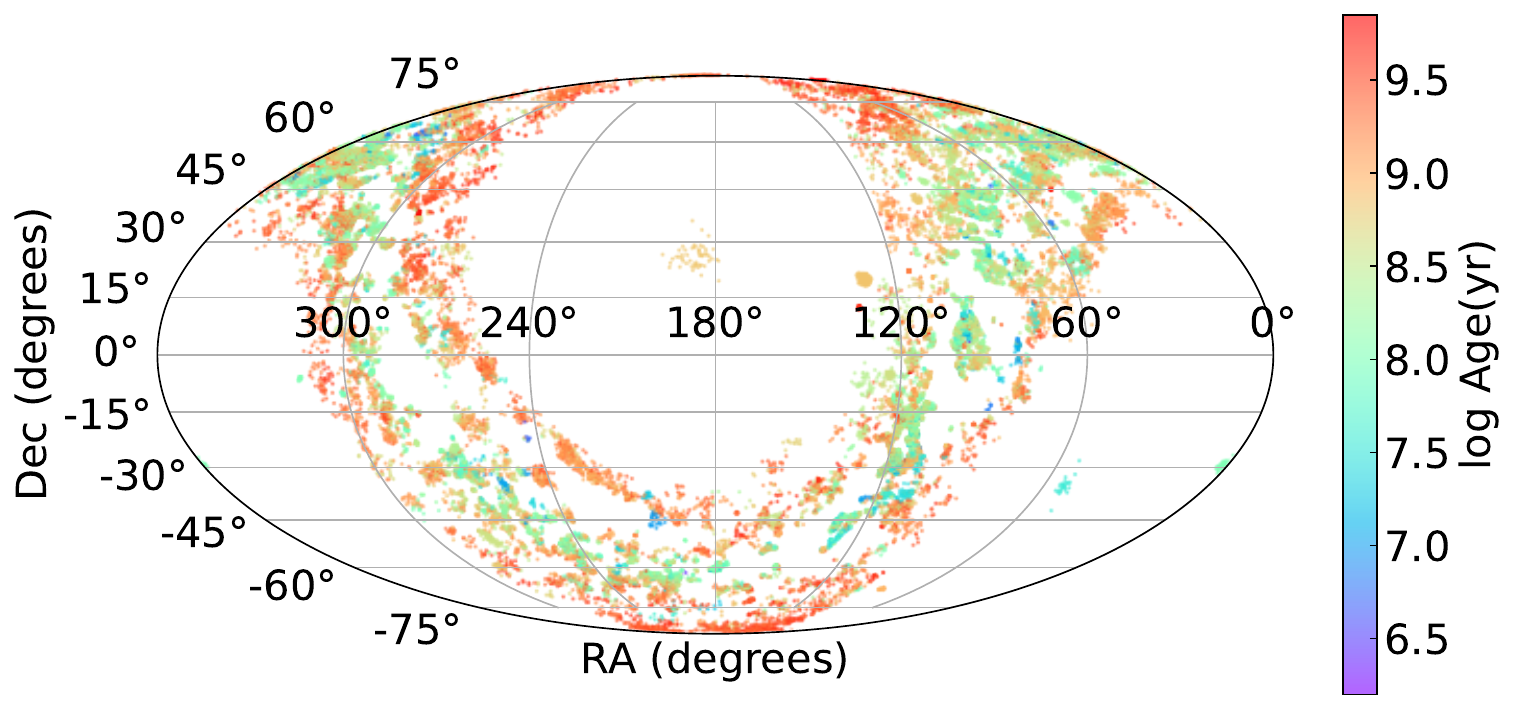}
    \label{hr_mollweide_oc.fig}}
    \subfigure[]{
    \includegraphics[width=0.46\textwidth]{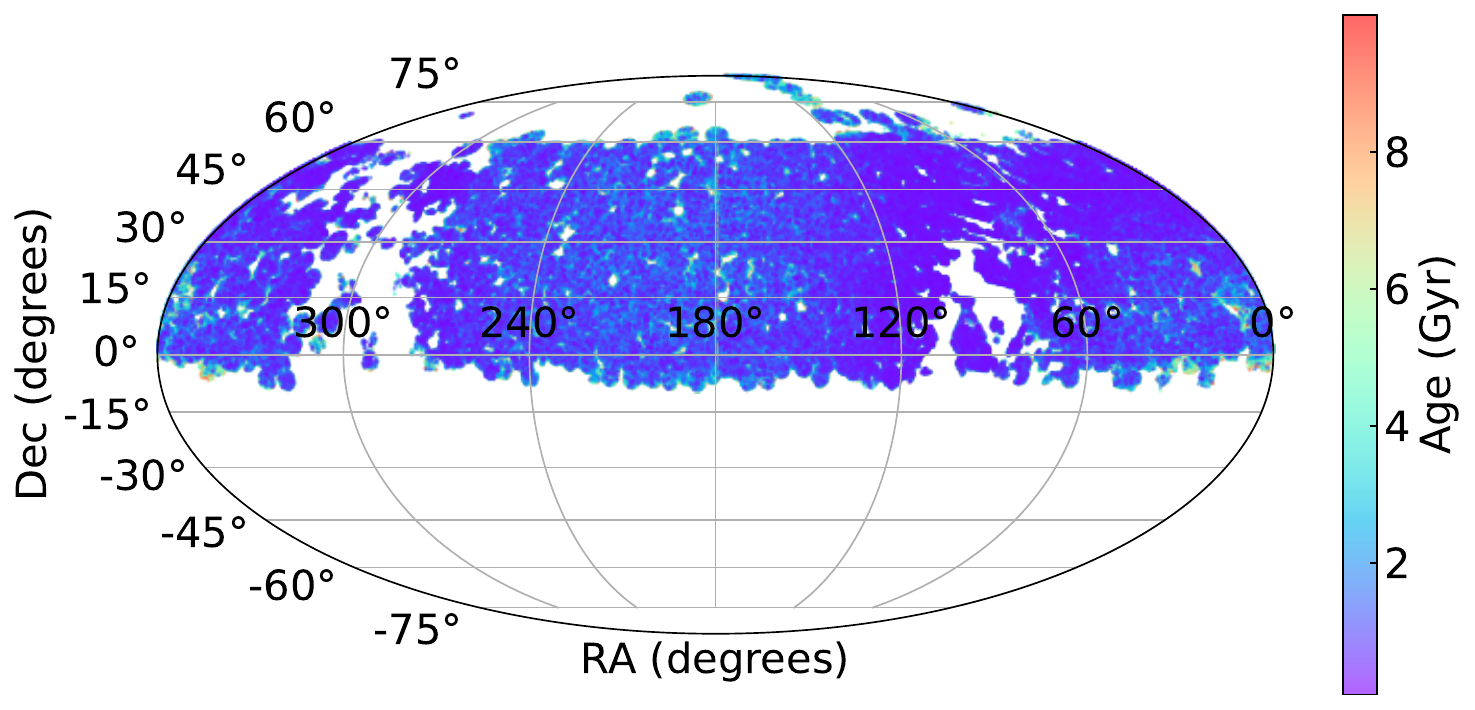}
    \label{hr_mollweide_field.fig}}
    \caption{Panel (a): H–R diagram of the open cluster members stars with GALEX observations. The blue solid line is the fitting results of main sequences. green dashed lines represent the upper and lower bounds of the main sequence fitted with a $\pm 0.5$ offset, and the red dashed lines is the same sequence shifted upwards by 2.
    Panel (b): H–R diagram of the field stars with GALEX observations. 
    Panel (c): The galactic distribution of the selected sources from open clusters. The color bar represent the age of the stars. 
    Panel (d): The galactic distribution of the selected sources from field stars.}
    \label{hr.fig}
\end{figure*}

\subsection{Sample construction}
\label{sample_selection.sec}

The GALEX is a NASA space-based ultraviolet telescope, conducted the first sky-wide ultraviolet (UV) surveys \citep{2007ApJS..173..682M}. It observed in two UV bands: far-ultraviolet (FUV; $\lambda_{\rm eff} \sim 1528$~\AA, $1344-1786$~\AA) and near-ultraviolet (NUV; $\lambda_{\rm eff} \sim 2310$~\AA, $1771-2831$~\AA). The most comprehensive release to date, GR6$+$7 \citep{2017ApJS..230...24B}, includes 82,992,086 unique sources observed through two major imaging programs: the All-Sky Imaging Survey (AIS; $\sim$100 s exposure) and the Medium-depth Imaging Survey (MIS; $\sim$1500 s exposure). And the $VisitPhotoObjAll$ catalog provides photometric measurements from individual visits, totaling 292,296,119 FUV and NUV detections. 
We obtained the GALEX data from the Mikulski Archive for Space Telescopes (MAST) \citep{https://doi.org/10.17909/t9h59d} and cross-matched our sample with the $VisitPhotoObjAll$ catalog using a 5$\arcsec$ radius through CasJobs\footnote{\url{https://mastweb.stsci.edu/kplrcasjobs/}}.

To resolve cases where multiple GALEX sources were found within the 5$\arcsec$ matching radius, we applied additional filtering: if their NUV magnitudes differ by more than 2.5 mag, we adopted the brightest source as the optical counterpart; if their magnitudes are similar, we selected the closest source as the true counterpart.
Furthermore, a single source may be observed multiple times by GALEX. For sources with multiple observations, we calculated the average magnitude, and obtained magnitudes for a total of 130,953 sources for open clusters and 2,379,826 sources for field stars.

We also calculated stellar absolute magnitudes from the Gaia Archive, in combination with the distances and extinction values $A_{V}$ to select main-sequence stars. 

For open clusters, we first excluded sources with ages older than 1~Gyr and $G_{\rm BP}-G_{\rm RP} < 0.5$ from the sample, which considering the old blue stars in this region of the Hertzsprung-Russell (HR) diagram are likely blue stragglers. Then we performed polynomial fitting to the data. This fitting was carried out iteratively 3 times to derive a range for the main sequence on the HR diagram. The main sequence was fitted as shown by the blue solid line in Figure \ref{hr_gaia_oc.fig}, and the fitting result is 

\begin{align}
\begin{split}
    M_{\rm G} = 0.20273\times(G_{\rm BP}-G_{\rm RP})^4\\ -1.11397\times(G_{\rm BP}-G_{\rm RP})^3\\ +1.30805\times(G_{\rm BP}-G_{\rm RP})^2\\ +3.96643\times(G_{\rm BP}-G_{\rm RP})\\ +0.81319.
\end{split}
\end{align}

We selected a range of $\pm$ 0.5 magnitudes around the fitted curve to define the main sequence (the green dashed lines in the Figure \ref{hr_gaia_oc.fig}). However, in clusters younger than 100~Myr, many K- and M-type stars are still in the pre-main-sequence phase and fall outside this range. Therefore, for clusters younger than 100~Myr, we extended the upper boundary of the main sequence by 2 magnitudes above the fitted curve to include more K- and M-type stars (as shown by the red dashed line in the Figure \ref{hr_gaia_oc.fig}). 
We finally only selected the stars with $E(B-V) < 0.5$ and distance $<5000$~pc, and obtained a final sample of 43,141 sources. The distribution of these sources in the celestial coordinate system is shown in Figure \ref{hr_mollweide_oc.fig}.
Their parameters are listed in Table \ref{uv_pars.tab}.

\begin{figure*}[!htbp]
    \centering
    \includegraphics[width=0.98\textwidth]{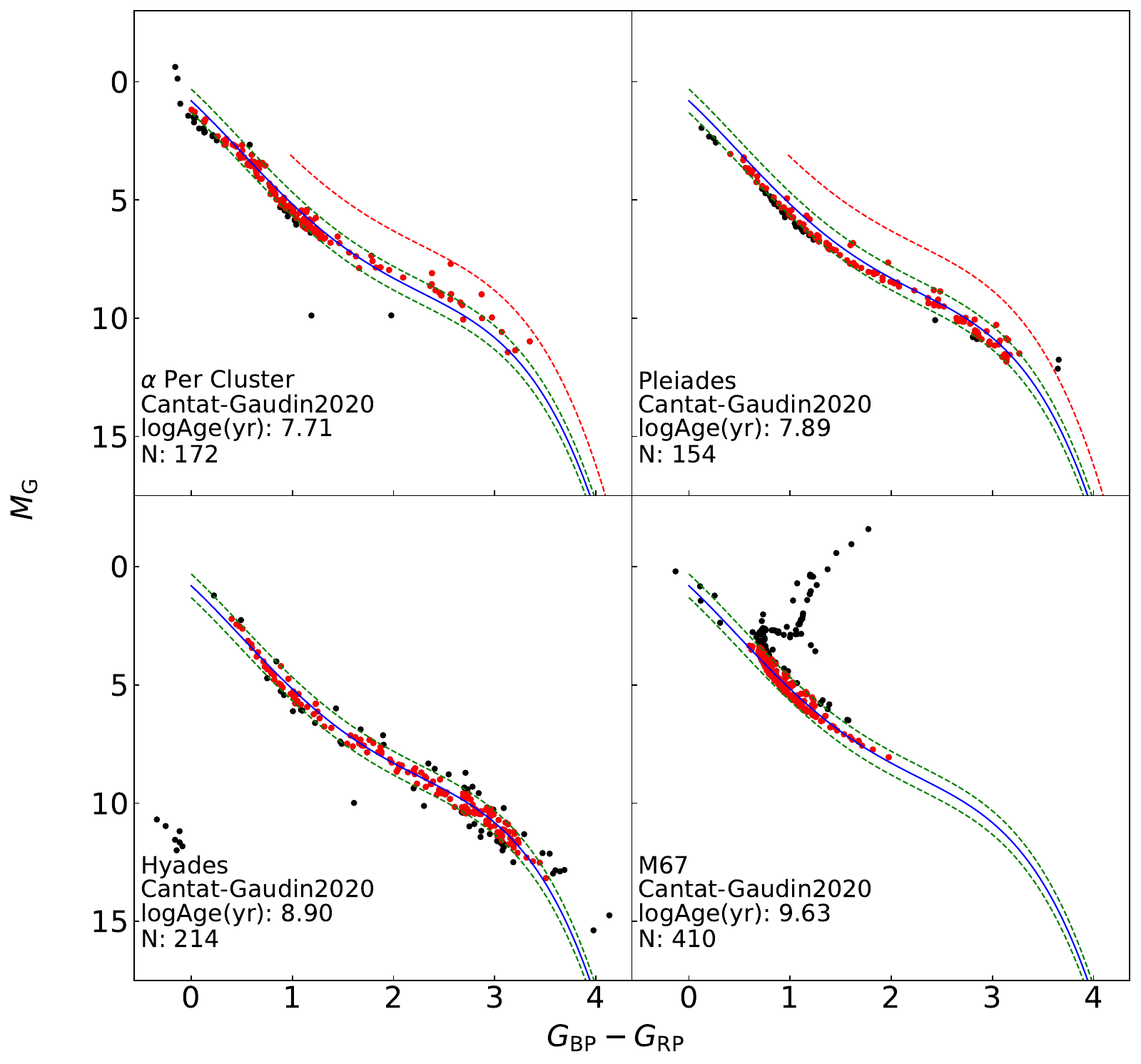}
    \caption{Clusters CMD of the $\alpha$ Per Cluster, Pleiades, Praesepe, Hyades, and M67. The black dots are the open cluster members with GALEX observations. The red dots are the main sequence samples that we defined. The different dash lines are the same as Figure \ref{hr.fig}.}
    \label{hr_4_clusters.fig}
\end{figure*}

Figure \ref{hr_4_clusters.fig} shows the member stars of four open clusters, $\alpha$ Per, Pleiades, Hyades, and M67 as examples. 
Comparing the HR diagrams before and after applying our selection criteria shows that we retain main-sequence stars and removes evolved ones, yielding a cleaner sample.

For field stars, we dropped the A-type stars ($G_{\rm BP}-G_{\rm RP}<0.37$) since it is hard to distinguish A-type main-sequence stars and blue stragglers. Then we selected the stars with the $E(B-V) < 0.5$ and distance $<5000$~pc, and performed polynomial fitting as same as the open clusters, and obtained the fitting result as,
\begin{align}
\begin{split}
    M_{\rm G} = 0.34062\times(G_{\rm BP}-G_{\rm RP})^4\\ -0.79000\times(G_{\rm BP}-G_{\rm RP})^3\\ -2.33483\times(G_{\rm BP}-G_{\rm RP})^2\\ +10.03523\times(G_{\rm BP}-G_{\rm RP})\\ +1.68015.
\end{split}
\end{align}
We selected a range of $\pm$ 0.5 magnitudes around the fitted curve as the range of the main sequence (the green dashed lines in the Figure \ref{hr_gaia_field.fig}). Finally, we obtained a main sequence catalog including 1,573,928 field stars.

\begin{figure*}[!htbp]
    \centering
    \subfigure[]{
    \includegraphics[width=0.47\textwidth]{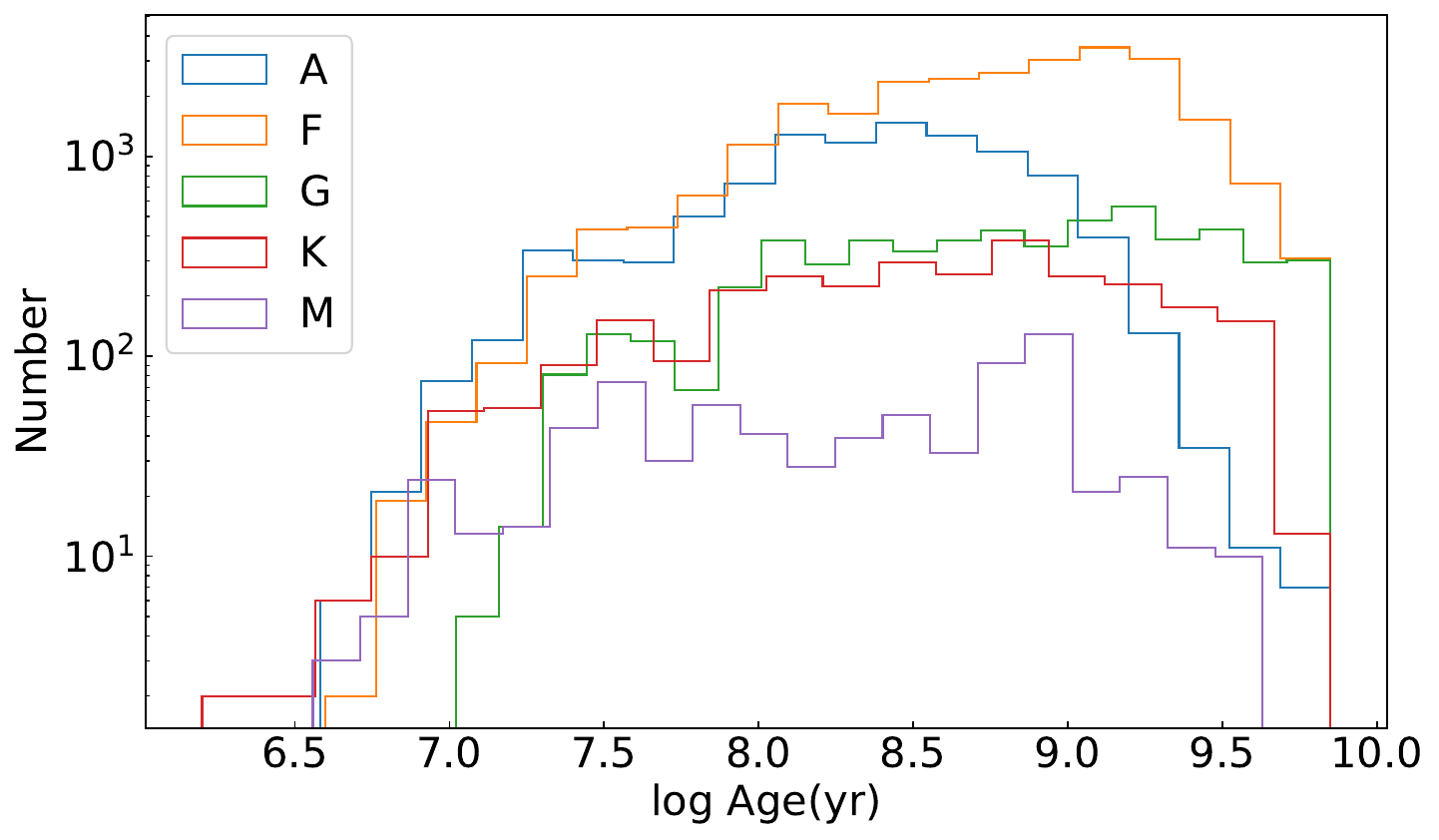}
    \label{logage_distribution.fig}}
    \subfigure[]{
    \includegraphics[width=0.47\textwidth]{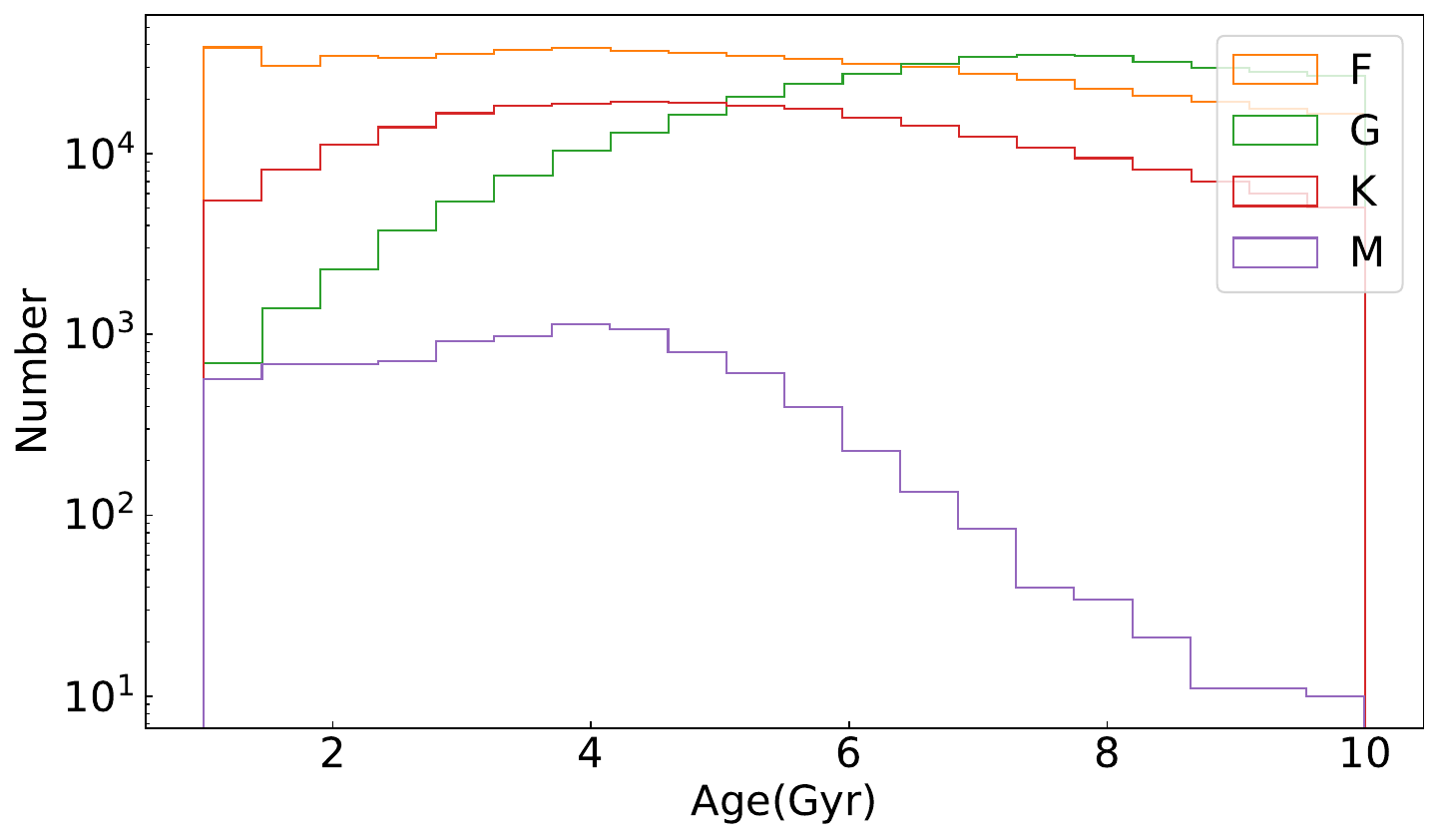}
    \label{age_distribution_field.fig}}
    \subfigure[]{
    \includegraphics[width=0.47\textwidth]{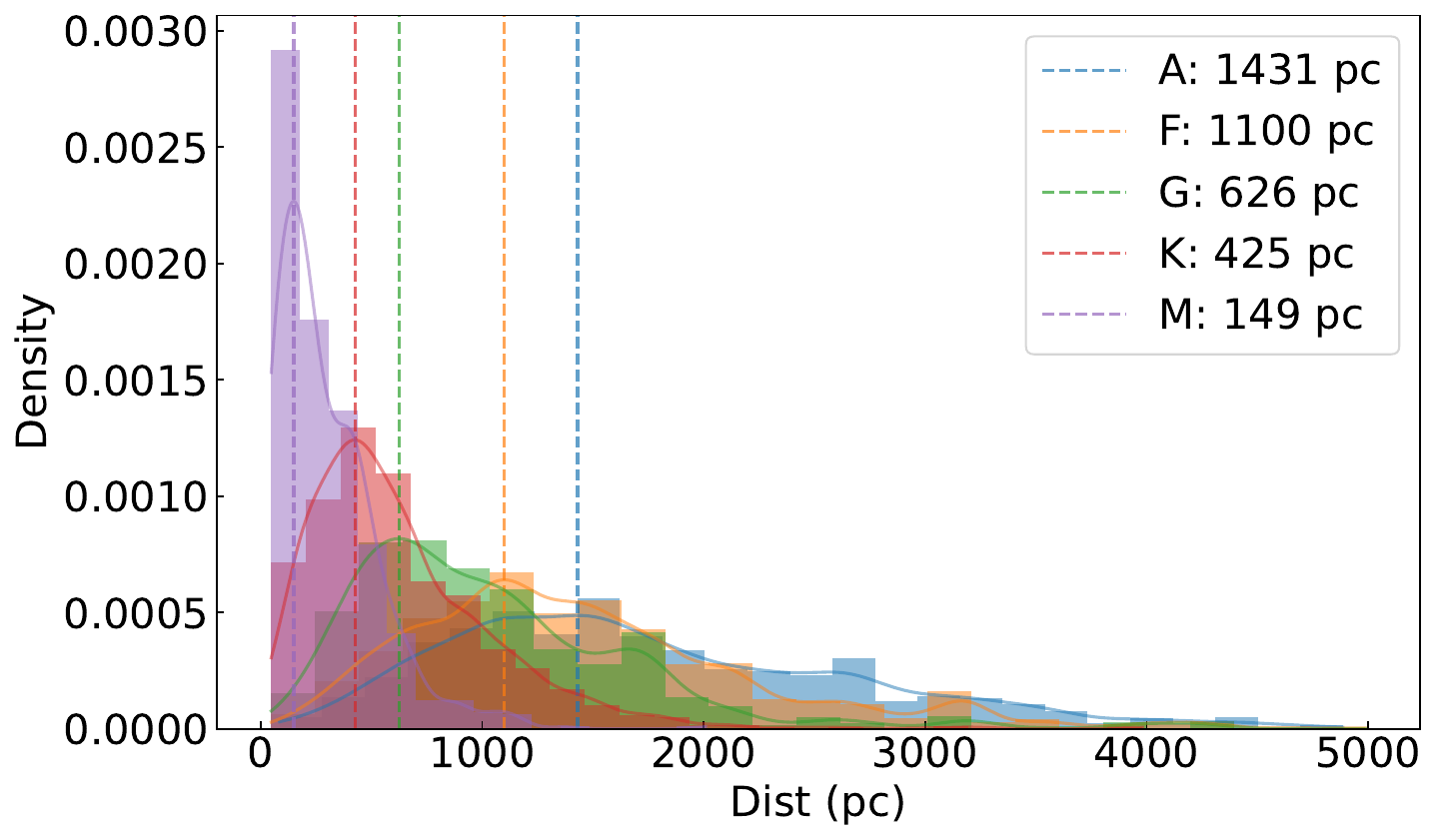}
    \label{dist_distribution.fig}}
    \subfigure[]{
    \includegraphics[width=0.47\textwidth]{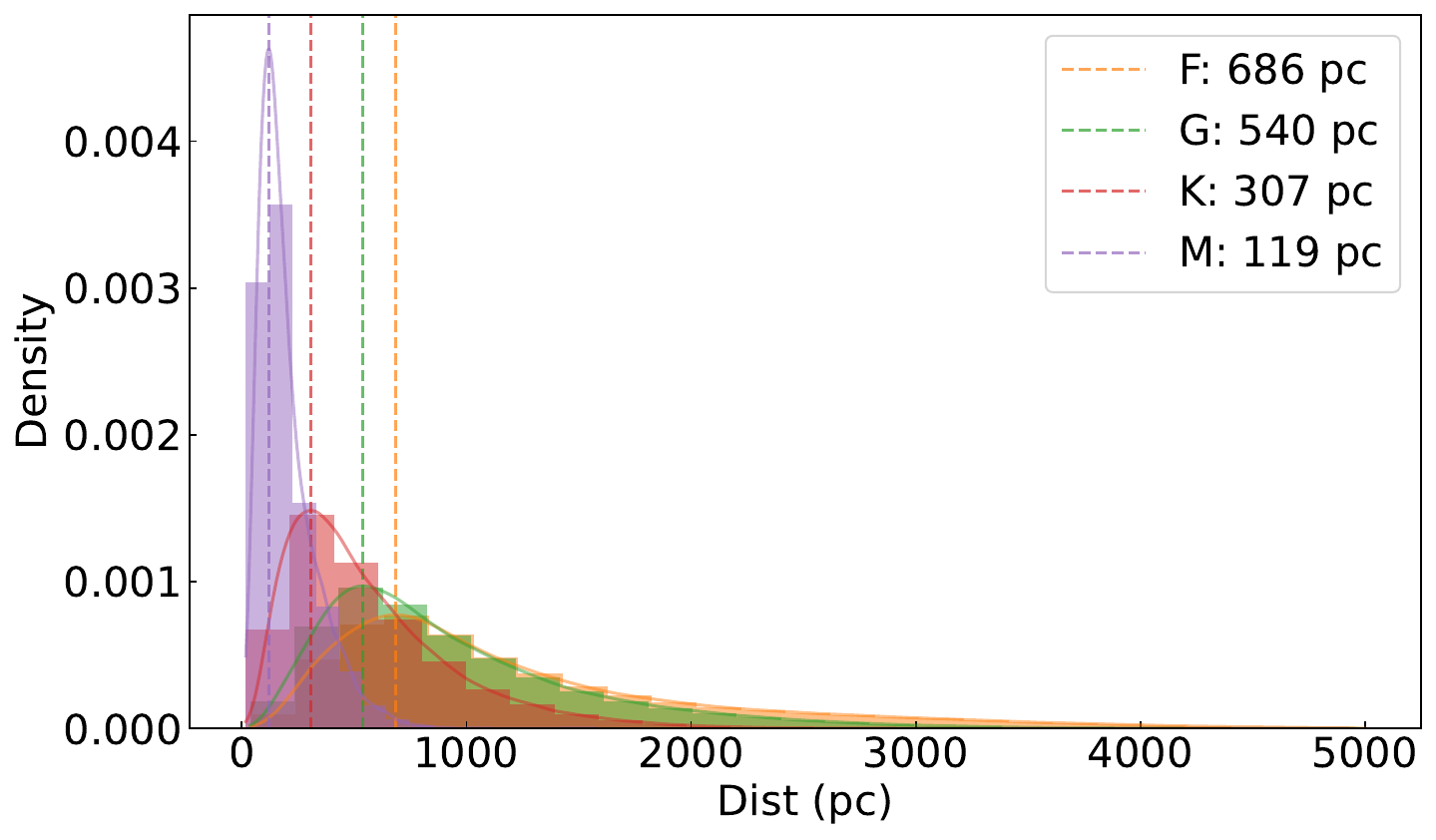}
    \label{dist_distribution_field.fig}}
    \caption{
    Panel (a): The distribution of age for the open clusters, and the blue, orange, green, pink, and purple lines represent the A, F, G, K, and M-type stars, respectively.
    Panel (b): The distribution of age for the field stars, and the different color lines are the same as Panel (a).
    Panel (c): The distribution of distance for the open clusters, and the blue, orange, green, pink, and purple bars are the A, F, G, K, and M-type stars, respectively. The solid lines with the same color are the fitting results, and the dashed lines are the peak density of different spectral type.
    Panel (d): The distribution of distance for the field stars. The different color lines are the same as Panel (c).
    }
    \label{pars_distribution.fig}
\end{figure*}

We divided our sample into A-M dwarf stars based on the $G_{\rm BP}-G_{\rm RP}$ values according to the Table 5 in \cite{2013ApJS..208....9P}. The distributions of their ages and distances are shown in Figure \ref{pars_distribution.fig}.

\section{Evolution of stllar NUV emission}
\label{uv_emission.sec}

We used $f_{\rm J}$ in $\rm \mu Jy$ to normalize the NUV emission for different type of stars. Because the $J$-band flux is primarily dominated by photospheric emission, and is less sensitive to stellar chromospheric and coronal activity.
Therefore, the NUV emission is characterized by the ratio $f_{\rm NUV}/f_{\rm J}$ \citep{2014AJ....148...64S}. We converted magnitudes to flux densities using the equation following \citep{2019ApJ...872...17R},

\begin{align}
\label{cps_flux.eq}
    f_{\rm NUV} \, (\rm \mu Jy) &= 10^{\frac{23.9-m_{\rm NUV}-R_{\rm NUV}\times E(B-V)}{2.5}}
\end{align}
and
\begin{align}
\label{cps_mag.eq}
    f_{\rm J} \, (\rm \mu Jy) &= 1.594\times 10^9 \times 10^{\frac{m_{\rm J}-R_{\rm J}\times E(B-V)}{-2.5}},
\end{align}
where $f_{\rm NUV}$ and $f_{\rm J}$ represent the flux density $(\rm \mu Jy$) in the NUV and J bands, respectively. $m_{\rm NUV}$ and $m_{\rm J}$ are the observed magnitudes in the NUV and J bands, respectively. $R_{\rm NUV}$ and $R_{\rm J}$ are the extinction coefficients. We used the values of 8.71 for NUV band, 8.11 for FUV band, and 0.7927 for J band, according to \cite{1989ApJ...345..245C}. The reddening $E(B-V)$ was calculated from $A_{V}/3.1$, assuming $R_{V} = 3.1$. For stars in the open clusters, we adopted the extinction values $A_{V}$ directly from the cluster catalogs, while for the field stars, we estimated the extinction using the $mwdust$ package \citep{2016ApJ...818..130B}, which provides the 3D dust reddening maps based on the Galactic coordinates and distances.
Photometric data in the NUV and FUV bands were obtained from the GALEX GR6+7 catalog (see Section~\ref{samples.sec}). We cross-matched these sources with the Two Micron All-Sky Survey (2MASS) using a $3\arcsec$ radius via TOPCAT \citep{2005ASPC..347...29T}, and retrieved J-band magnitudes. The calculated parameters are listed in Table \ref{uv_results.tab}.

Considering the completeness of the sample, we fitted the distance distribution of the sample using Guassian kernel density estimation and selected only sources within the peak distance for each spectral type for following analyses. For the open clusters (Figure \ref{dist_distribution.fig}), the completeness-limited distances are 1,431, 1,100, 626, 425, and 149 pc for A-, F-, G-, K-, and M-type dwarfs, respectively. For the field stars (Figure \ref{dist_distribution_field.fig}), they are closer, with the distances of 686, 540, 307, and 121 pc for F-, G-, K-, and M-dwarfs, respectively. 

\begin{table*}[htp]
\centering
\caption{The basic parameters of our sample.}
\label{uv_pars.tab}
\resizebox{\textwidth}{!}{
\begin{threeparttable}
    \begin{tabular}{l*{8}{c}}
\hline
Gaia DR3 ID & Cluster & Member Number & log Age & Dist & $A_{V}$ & Ref.\tnote{1} & R.A.(J2000) & Decl.(J2000) \\
 &  &   &   &  (pc) & (mag) &   &  ($^\circ$) &  ($^\circ$) \\
\hline
39356988653848576 &             \text{Theia\_7135} &          48 &    9.17 &  1152 & 1.43 &   K20 & 60.081179 &  15.124635 \\
39370114074031488 &             \text{Theia\_7135} &          48 &    9.17 &  1152 & 1.43 &   K20 & 60.171416 &  15.354407 \\
39388977570246656 &             \text{Theia\_7135} &          48 &    9.17 &  1152 & 1.43 &   K20 & 60.271772 &  15.269348 \\
39406226158867968 &             \text{Theia\_7135} &          48 &    9.17 &  1152 & 1.43 &   K20 & 60.448917 &  15.659178 \\
40057789877836032 &             \text{Theia\_7135} &          48 &    9.17 &  1152 & 1.43 &   K20 & 58.516633 &  14.995120 \\
40058339633644928 &             \text{Theia\_7135} &          48 &    9.17 &  1152 & 1.43 &   K20 & 58.553716 &  15.025949 \\
42716408993806080 & \text{Melotte\_25/Theia\_1004} &         495 &    8.90 &    47 & 0.00 & Can20 & 53.143707 &  16.153328 \\
43538293935879680 & \text{Melotte\_25/Theia\_1004} &         495 &    8.90 &    47 & 0.00 & Can20 & 58.777088 &  16.998478 \\
43575746050631808 &             \text{Theia\_6030} &          63 &    8.93 &   885 & 1.36 &   K20 & 58.193495 &  17.100379 \\
43789772861265792 & \text{Melotte\_25/Theia\_1004} &         495 &    8.90 &    47 & 0.00 & Can20 & 57.604027 &  17.246509 \\
44052178182130816 & \text{Melotte\_25/Theia\_1004} &         495 &    8.90 &    47 & 0.00 & Can20 & 55.518474 &  16.421759 \\
44894472808190720 & \text{Melotte\_25/Theia\_1004} &         495 &    8.90 &    47 & 0.00 & Can20 & 55.481208 &  18.759936 \\
45142206521351552 & \text{Melotte\_25/Theia\_1004} &         495 &    8.90 &    47 & 0.00 & Can20 & 61.924932 &  15.162790 \\
45198178534988672 & \text{Melotte\_25/Theia\_1004} &         495 &    8.90 &    47 & 0.00 & Can20 & 62.489853 &  15.416981 \\
45293526808851200 & \text{Melotte\_25/Theia\_1004} &         495 &    8.90 &    47 & 0.00 & Can20 & 62.865194 &  15.992152 \\
\hline
\end{tabular}
    \begin{tablenotes}
        \normalsize 
        \item[1] Can20, Cas20, K20, and H24 represent the \cite{2020A&A...633A..99C},  \cite{2020A&A...635A..45C},  \cite{2020AJ....160..279K},  \cite{2024A&A...686A..42H}, respectively. 
    \end{tablenotes}
    \end{threeparttable}
}
\end{table*}

\begin{table*}[htp]
\centering
\caption{The calculated results of our sample.}
\label{uv_results.tab}
\resizebox{\textwidth}{!}{
\begin{threeparttable}
    \begin{tabular}{l*{6}{c}}
\hline
    Gaia DR3 ID & $G_{\rm BP}-G_{\rm RP}$ & $M_{\rm G}$ &  $m_{\rm NUV}$ & Jmag  & $f_{\rm NUV}$/$f_{\rm J}$& log$L_{\rm NUV}$ \\
     &  (mag) & (mag) & (mag) & (mag) & & log$\rm (erg/s)$  \\
\hline
39356988653848576 &   0.60 &  2.94 &  21.43 $\pm$ 0.1 & 12.95 $\pm$ 0.02 &              -1.57 &     32.61 \\
39370114074031488 &   0.71 &  4.34 & 23.13 $\pm$ 0.46 & 14.27 $\pm$ 0.03 &              -1.72 &     31.93 \\
39388977570246656 &   0.62 &  3.47 & 21.37 $\pm$ 0.35 & 13.45 $\pm$ 0.02 &              -1.35 &     32.63 \\
39406226158867968 &   0.79 &  4.10 & 22.86 $\pm$ 0.04 & 13.89 $\pm$ 0.02 &              -1.77 &     32.04 \\
40057789877836032 &   0.78 &  4.66 & 21.88 $\pm$ 0.49 & 14.48 $\pm$ 0.03 &              -1.14 &     32.43 \\
40058339633644928 &   0.56 &  3.28 & 21.01 $\pm$ 0.28 & 13.32 $\pm$ 0.02 &              -1.26 &     32.77 \\
42716408993806080 &   2.74 &  9.60 & 21.42 $\pm$ 0.45 &  10.1 $\pm$ 0.02 &              -4.17 &     28.23 \\
43538293935879680 &   1.06 &  5.38 & 15.86 $\pm$ 0.03 &  7.41 $\pm$ 0.02 &              -3.02 &     30.45 \\
43575746050631808 &   0.68 &  4.25 & 21.85 $\pm$ 0.41 & 13.53 $\pm$ 0.03 &              -1.58 &     32.13 \\
43789772861265792 &   1.08 &  5.85 &  13.75 $\pm$ 0.0 &  7.78 $\pm$ 0.02 &              -2.03 &     31.29 \\
44052178182130816 &   3.15 & 11.48 & 21.38 $\pm$ 0.33 & 11.61 $\pm$ 0.02 &              -3.55 &     28.24 \\
44894472808190720 &   2.31 &  8.87 & 20.94 $\pm$ 0.22 &  9.71 $\pm$ 0.02 &              -4.14 &     28.42 \\
45142206521351552 &   0.51 &  2.62 &  12.07 $\pm$ 0.0 &  5.24 $\pm$ 8.89 &              -2.37 &     31.96 \\
45198178534988672 &   3.14 & 11.22 &  22.32 $\pm$ 0.4 & 11.37 $\pm$ 0.03 &              -4.02 &     27.86 \\
45293526808851200 &   2.82 & 10.42 & 20.93 $\pm$ 0.18 & 10.81 $\pm$ 0.02 &              -3.69 &     28.42 \\
\hline
\end{tabular}
    \end{threeparttable}
}
\end{table*}

\subsection{NUV emission across different types of dwarfs}
\label{uv_type.sec}
 
\begin{figure*}[!htbp]
\centering
    \includegraphics[width=0.98\textwidth]{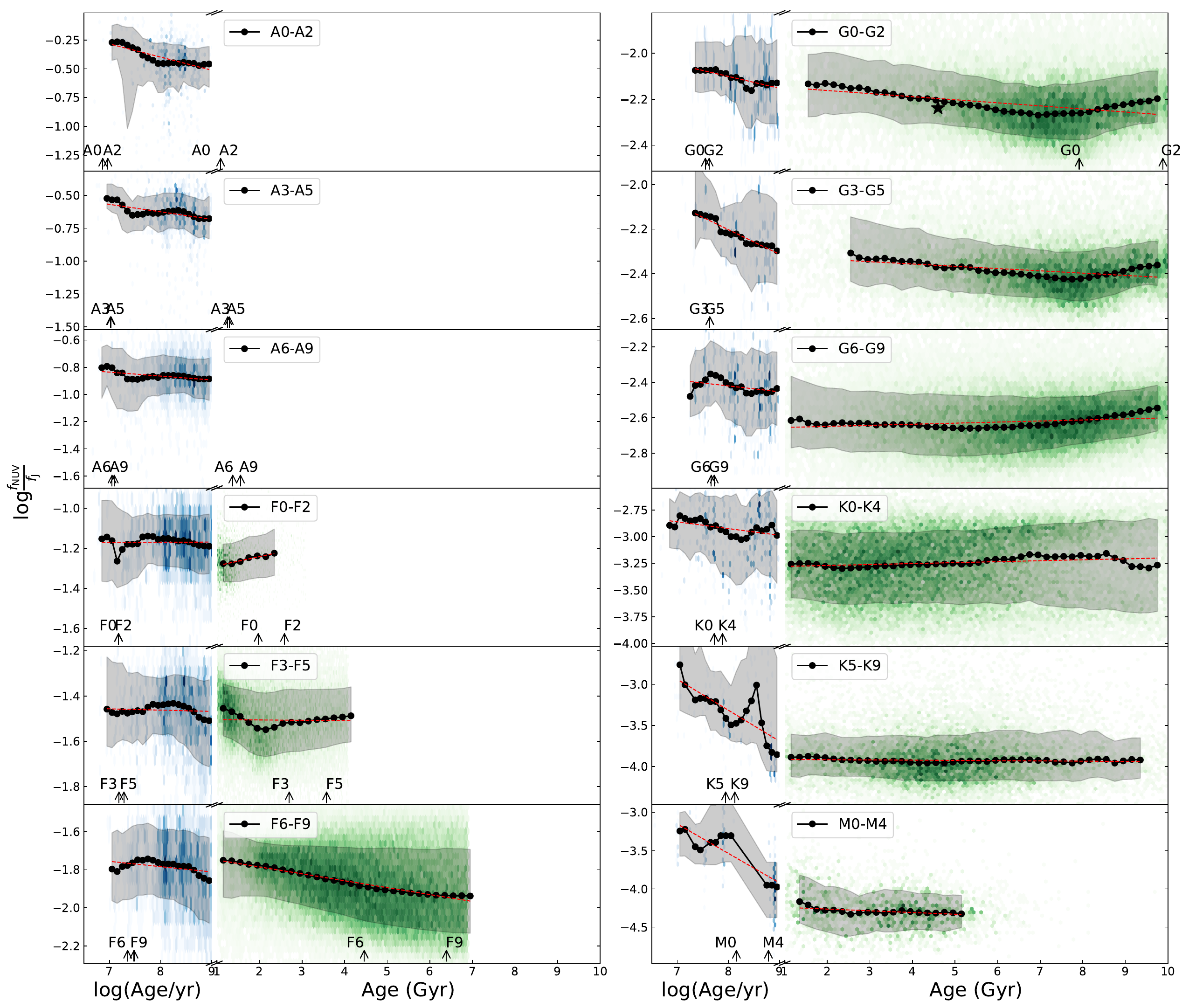}
    \caption{The relation between the log$f_{\rm NUV}/f_{\rm J}$ and stellar age considering the distance completeness. The colored points are number density. The blue points and the green points represent the open clusters and the field stars, respectively. The black dots are the median value of each bin for age, and the gray shaded regions are the 16\% to 84\% of each bin. The red dotted line is the result of a linear fit using black points. The gray arrows indicate the begin and end times of the main-sequence phase for different subtypes. The black star is the Sun.}
    \label{distribution_fnuv_oc_field.fig}
\end{figure*}

\begin{figure*}[!htbp]
\centering
    \includegraphics[width=0.98\textwidth]{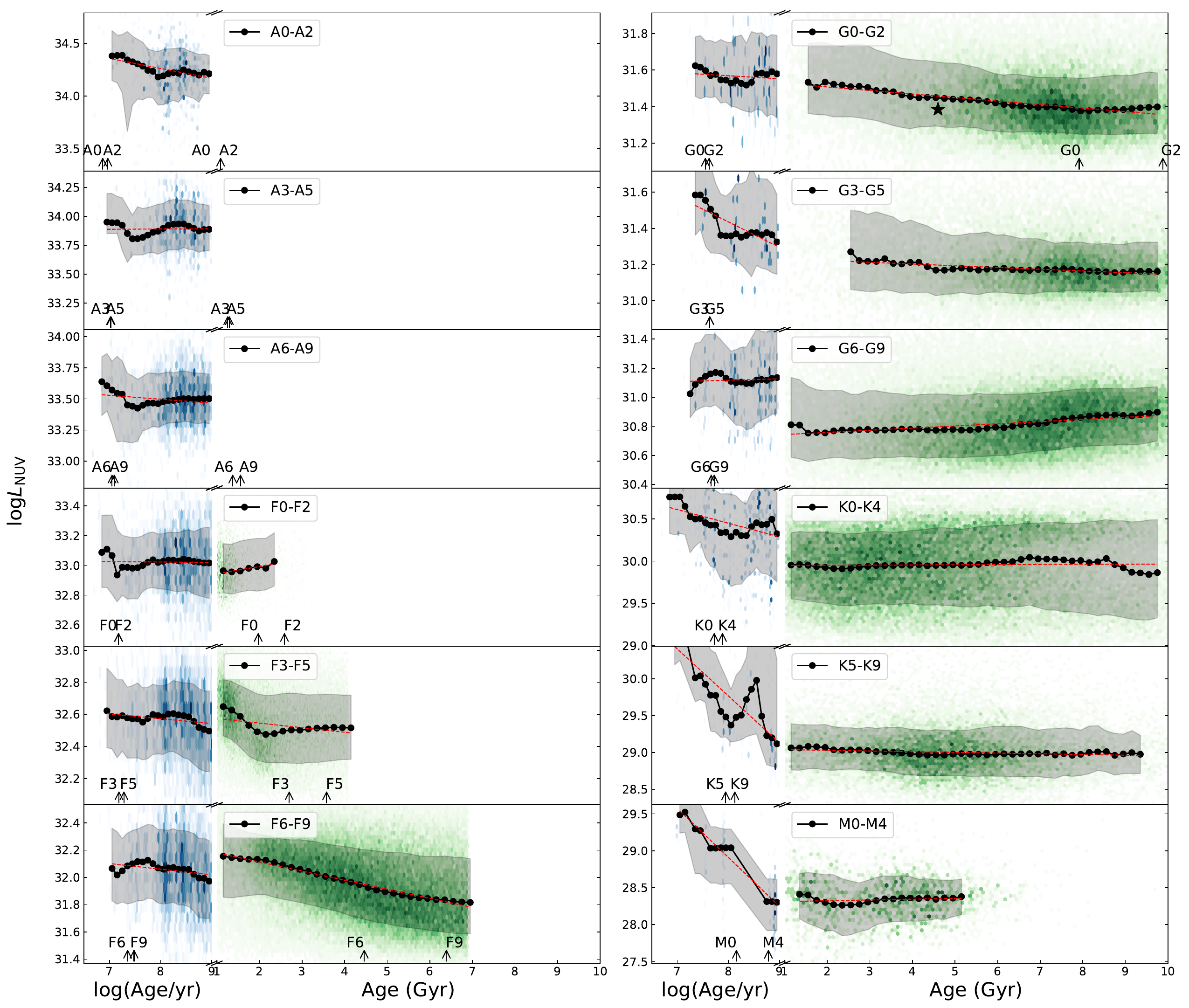}
    \caption{The relation between the log$L_{\rm NUV}$ and stellar age considering the distance completeness. The symbols are the same as the Figure \ref{distribution_fnuv_oc_field.fig}.}
    \label{distribution_lnuv_oc_field.fig}
\end{figure*}

\begin{figure*}[!htbp]
\centering
    \includegraphics[width=0.98\textwidth]{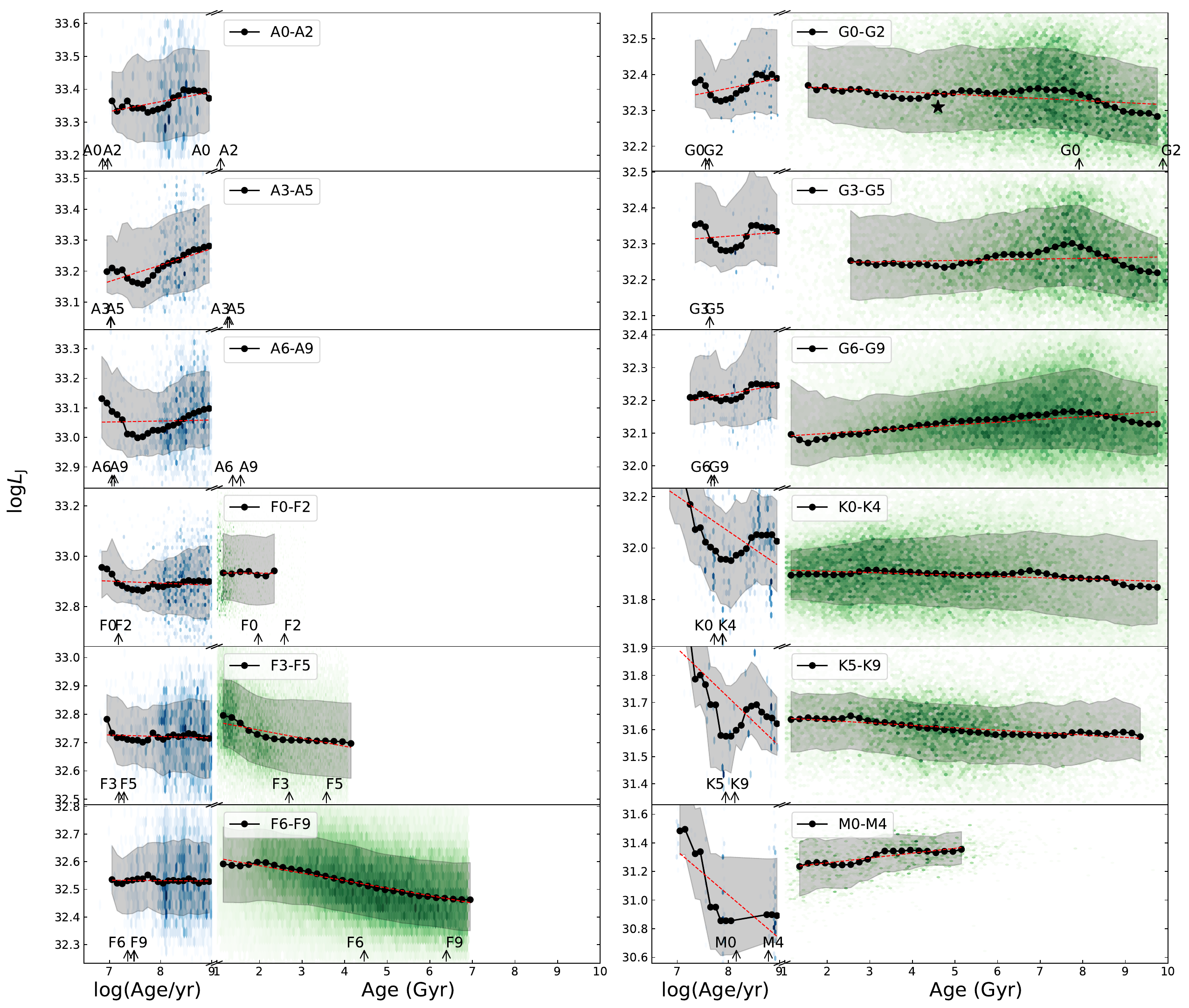}
    \caption{The relation between the log$L_{\rm J}$ and stellar age considering the distance completeness. The symbols are the same as Figure \ref{distribution_fnuv_oc_field.fig}.}
    \label{distribution_lj_oc_field.fig}
\end{figure*}

We analyzed the relationship between NUV emission and stellar age for different spectral types, with ages ranging from a few hundred million years to ten billion years. 
To better capture the evolutionary trends, we applied a moving median filter method. For the open cluster stars, starting from a logarithmic age of 6.5, we constructed age bins with an initial step size of 0.5~dex. Then, for each successive bin, the step size was gradually increased by 0.1~dex to generate a moving median sequence, continuing until the age reached 1~Gyr. For the field stars, we adopted a linear age scale from 1~Gyr to 10~Gyr, using bins of 0.5~Gyr in width, with a step size of 0.2~Gyr for the moving window. 
In addition, we classified the stars by spectral sub-type (from A0 to M4). We also identified the ages at which stars of different sub-types enter and leave the main sequence based on the MIST (MESA Isochrones and Stellar Tracks) EEP (Evolutionary Extinction Packagemodel)\footnote{\url{https://waps.cfa.harvard.edu/MIST/model_grids.html\#eeps}} \citep{2011ApJS..192....3P, 2016ApJ...823..102C, 2016ApJS..222....8D}, and these transition points are marked in Figure \ref{distribution_fnuv_oc_field.fig}. 
As our analysis focuses on main-sequence behavior, we excluded stars that have evolved beyond the main sequence from our sample. 
The evolution of NUV emission ($f_{\rm NUV}/f_{\rm J}$ and $L_{\rm NUV}$) with age for different stellar types are presented in Figures \ref{distribution_fnuv_oc_field.fig} and  \ref{distribution_lnuv_oc_field.fig}, as well as the evolution of J-band emission shown in Figure \ref{distribution_lj_oc_field.fig}. The NUV emission exhibits a clear evolutionary trend across different stellar types, with notable variations depending on spectral type . 

For A-type and early F-type stars, NUV emission is primarily driven by blackbody radiation due to their relatively high temperatures, along with hydrogen ionization and metal line emissions from ionized metals. As these stars evolve rapidly, their radii expand, leading to a decrease in temperature and an increase of $J$-band emission (Figure \ref{distribution_lj_oc_field.fig}), which in turn results in a gradual weak decline in NUV emission. This decrease is most pronounced in early A-type stars (e.g., A0-A2) (Figure \ref{distribution_fnuv_oc_field.fig}).

For late F-type to M-type stars, their NUV emission is primarily linked to magnetic activity.
As shown by \cite{2024ApJ...976...43W}, the contribution of magnetic activity to the observed NUV flux is about $15\%$ to $70\%$ for F and G stars, around $30\%$-$90\%$ for K stars, and more than $80\%$ for M stars.
Several phenomena can be seen in Figures \ref{distribution_fnuv_oc_field.fig} and  \ref{distribution_lnuv_oc_field.fig}.
First, a clear decrease in NUV emission is observed for late F- to G-type stars after 1~Gyr, indicating continuous spin-down due to magnetic braking over long-term stellar evolution \citep{2013ApJ...776...67V, 2016Natur.529..181V}.
Second, there is a sharp decline in NUV emission for late K- to M-type stars from 7~Myr to 1~Gyr, similar to trends observed in previous studies of M dwarfs \citep[e.g.,][]{2014AJ....148...64S, 2020ApJ...895....5P}. This decline occurs as stars evolve from young stellar objects (YSOs) to main-sequence stars. During the YSO stage, rapid rotation and the interaction between stars and surrounding disks result in high magnetic activity and thus enhanced NUV emission.
Third, the NUV emission for K- and M-type stars after 1~Gyr remains nearly stable \citep[e.g.,][]{2019ApJ...872...17R, 2023ApJ...951...44R}, suggesting slow rotational evolution. Cool stars exhibit much slower rotational evolution compared to their hotter counterparts during the main-sequence phase.
Fourth, we observe slight increases in NUV emission for older G- and early K-type stars. 
Finally, the Sun lies near the median NUV emission level for G-type dwarfs, suggesting that it is quite possible to find a ``Sun 2.0" among solar-like stars with similar NUV emission levels.

The underlying cause of the enhanced NUV emission observed in older G- and early K-type stars remains uncertain, but it may be linked to contamination from undetected white dwarf companions, often referred to as ``blue lurkers'' \citep{2019ApJ...881...47L}. The absence of a similar increase in late K and M stars can be explained. First, solar-type stars exhibit higher multiplicity fractions than late K and M dwarfs \citep[e.g.,][]{2010ApJS..190....1R, 2021ApJ...922..211N}. Second, all old field stars in our sample have LAMOST spectra. White dwarf companions to M dwarfs can be easily identified in the spectra, ensuring such systems are not classified as normal stars. However, detecting white dwarf companions to the brighter G- and early K-type stars is much more challenging.

Previous studies \citep[e.g.,][]{2014AJ....148...64S, 2018AJ....155..122S, 2019ApJ...872...17R} have also investigated the evolution of stellar NUV emission using samples of young moving groups, clusters aged between 10 and 625~Myr, and field stars around 5~Gyr. These studies similarly found that NUV emission from late-K and M dwarfs declines with age. However, unlike our results, their analysis, based on median UV flux evolution while accounting for stellar lifetimes, showed that the NUV emission continues to decline in older stars. One possible reason is that these studies included only a limited field sample of old K dwarfs (within 30 pc) and M dwarfs (within 10 pc), all of which were assigned a uniform age of 5 Gyr, which may not reflect their true ages. A test using K dwarfs within 30 pc from \cite{2025ApJS..280...13W} shows that these stars span a broad age range, primarily between 1 and 5 Gyr.

Stellar activity variability can influence the measured NUV flux, particularly for late-type dwarfs. For instance, large-amplitude flares can temporarily boost stellar UV flux, and the mean value derived from multiple observations may therefore be biased. Such effects may partly account for the scatter seen in Figure \ref{distribution_fnuv_oc_field.fig}.


The evolutionary trend of $L_{\rm NUV}$ (shown in Figure \ref{distribution_lnuv_oc_field.fig}) closely resembles that of $f_{\rm NUV}/f_{\rm J}$.
It is interesting to note that for M dwarfs, the $J$-band emission shows a sharper decline than the NUV emission before 100~Myr (as shown in Figure \ref{distribution_lj_oc_field.fig}), while stars older than 1~Gyr show much higher $J$-band emission than that of younger stars.
We also showed the NUV emission evolution and $J$-band luminosity without considering the distance completeness in Figure \ref{distribution_fnuv_all.fig}, \ref{distribution_lnuv_all.fig}, and \ref{distribution_lj_all.fig}.

\begin{table}[]
    \centering
    \caption{Slope of linear fit Figures \ref{distribution_fnuv_oc_field.fig} and \ref{distribution_lnuv_oc_field.fig}.}
    \label{slope.tab}
    \begin{tabular}{crrrr}
    \hline
         Stellar Type & \multicolumn{2}{c}{$f_{\rm NUV}/f_{\rm J}$} & \multicolumn{2}{c}{$L_{\rm NUV}$} \\
          & OC & Field & OC & Field \\
          \hline
        A0-A2 & $-0.11$ &  & $-0.09$ & \\
        A3-A5 & $-0.06$ &  & 0.00 & \\
        A6-A9 & $-0.03$ &  & $-0.03$ & \\
        F0-F2 & 0.00 & 0.05 & 0.00 & 0.05 \\
        F3-F5 & $-0.01$ & 0.00 & $-0.03$ & $-0.03$ \\
        F6-F9 & $-0.03$ & $-0.04$ & $-0.04$ & $-0.07$ \\
        G0-G2 & $-0.05$ & $-0.01$ & $-0.02$ & $-0.02$ \\
        G3-G5 & $-0.11$ & $-0.01$ & $-0.14$ & -0.01 \\
        G6-G9 & $-0.03$ & 0.01 & 0.01 & 0.01 \\
        K0-K4 & $-0.06$ & 0.01 & $-0.16$ & 0.00 \\
        K5-K9 & $-0.38$ & 0.00 & $-0.64$ & $-0.01$ \\
        M0-M4 & $-0.38$ & $-0.02$ & $-0.66$ & 0.01 \\
         \hline
    \end{tabular}
    NOTE. The OC and Field represent the open clusters and the field stars, respectively.
\end{table}

\begin{figure*}[!htbp]
    \centering
    \subfigure[]{
    \includegraphics[width=0.45\textwidth]{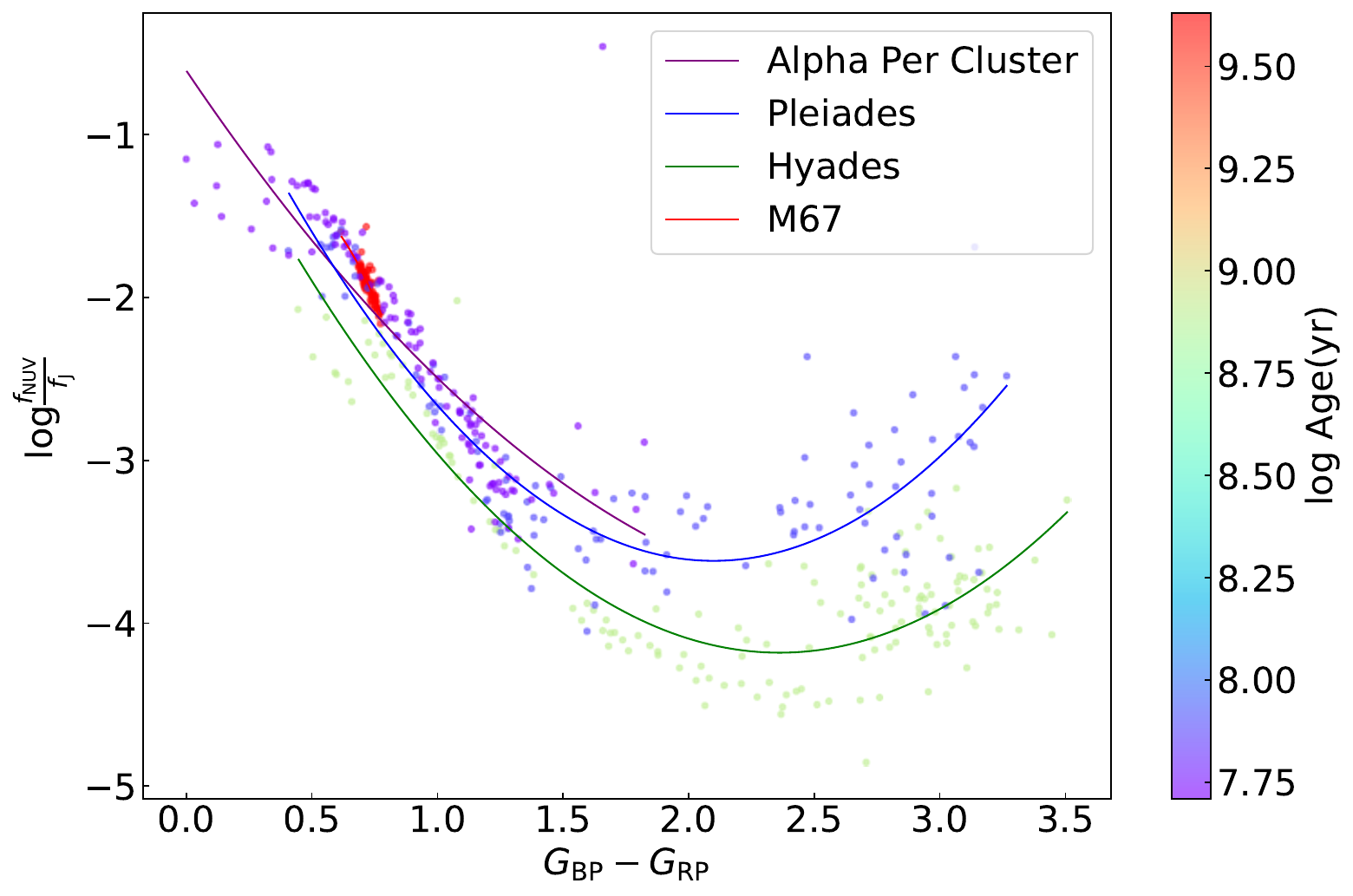}
    \label{fnuvfj_logage_dist_4.fig}}
    \subfigure[]{
    \includegraphics[width=0.45\textwidth]{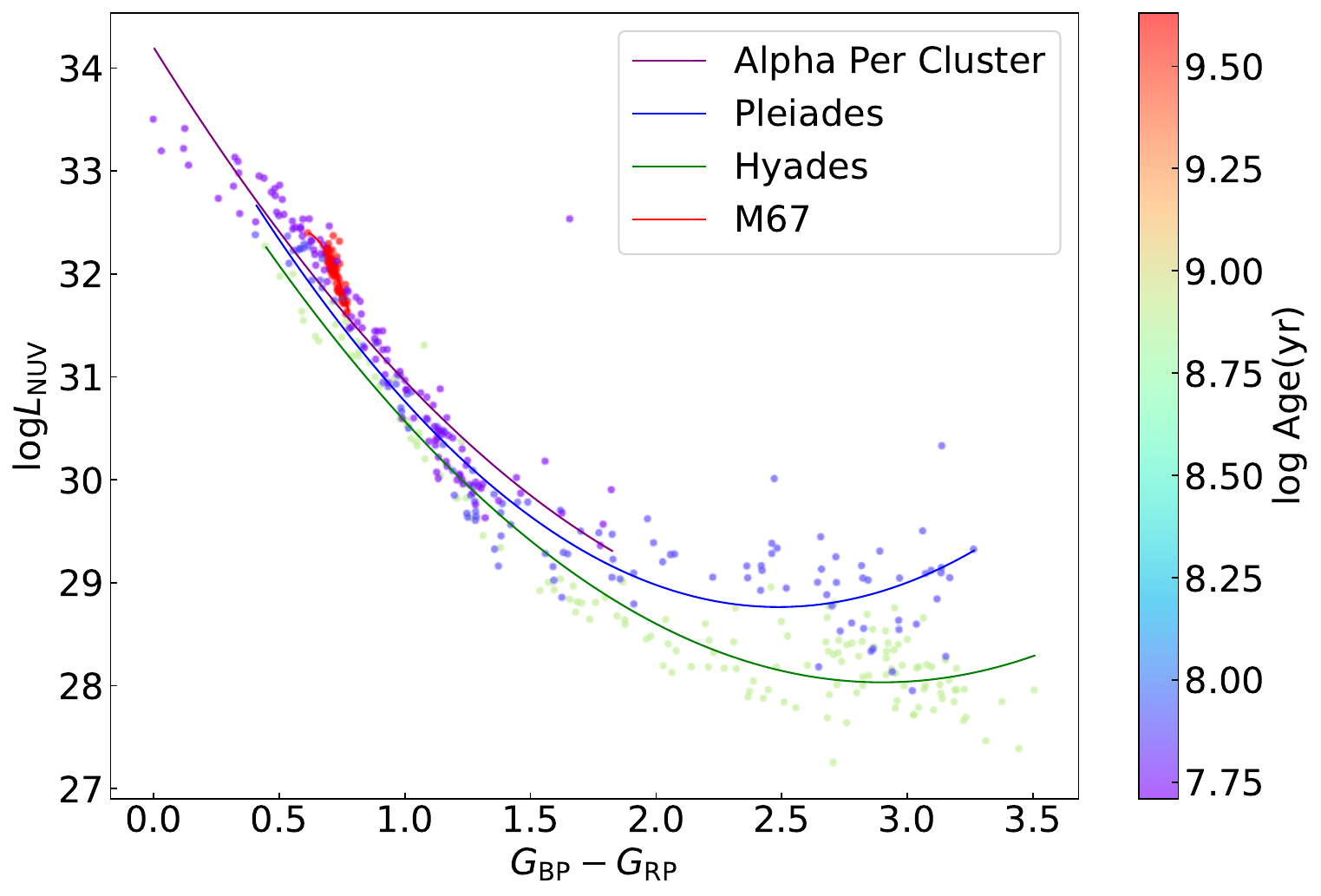}
    \label{lnuv_logage_dist_4.fig}}
    \caption{
    Panel (a): Scatter of log$f_{\rm NUV}/f_{\rm J}$ versus $G_{\rm BP}-G_{\rm RP}$ for $\alpha$ Per Cluster, Pleiades, Hyades, and M67. The color bar represents logAge in years, ranging from blue (younger) to red (older). The colored lines are the result of a polynomial fit for different star clusters.
    Panel (b): The same as Figure \ref{fnuvfj_logage_dist_4.fig}, but $y$ axis is the log$L_{\rm NUV}$.}
    \label{nuv_hr.fig}
\end{figure*}

We also examined the relationship between NUV emission and color for several open clusters, i.e., $\alpha$ Per, Pleiades, Hyades, and M67. As shown in Figure \ref{nuv_hr.fig}, for F-, G-, and K-type stars ($G_{\rm BP}-G_{\rm RP} < 1.5$), the values of log($f_{\rm NUV}/f_{\rm J}$) and log$L_{\rm NUV}$ vary only slightly across clusters of different ages. In contrast, M-type stars exhibit significant variation in both log($f_{\rm NUV}/f_{\rm J}$) and log$L_{\rm NUV}$. This is likely due to the presence of many pre-main-sequence M-type stars in our open cluster sample, which tend to rotate more rapidly and display stronger magnetic activity, leading to enhanced NUV emission in young stars compared to their old counterparts.

\subsection{Impact of metallicity on NUV emission}
\label{feh.sec}

\begin{figure*}[!htbp]
\centering
    \includegraphics[width=0.98\textwidth]{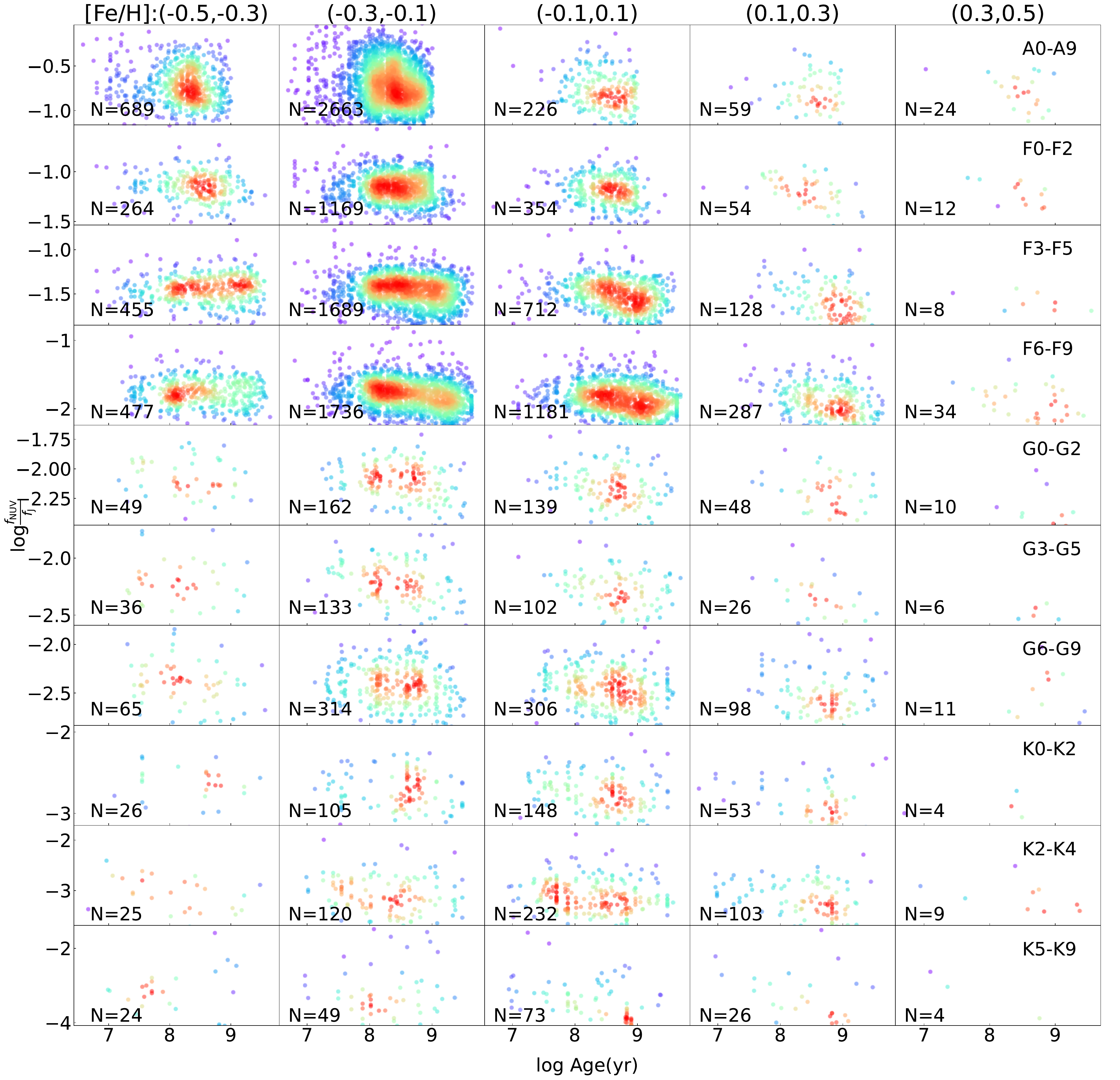}
    \caption{The relationship between NUV emission and stellar age for open clusters ranging from spectral type A to M-dwarfs. The colored points are number density. The black dots and the gray shaded regions correspond to the same as shown in Figure \ref{distribution_fnuv_oc_field.fig}. The red and blue dots also represent the same as the black dots. From left to right panel, the columns represent samples with metallicities $-0.5<\rm{[Fe/H]}<-0.3$, $-0.3<\rm{[Fe/H]}<-0.1$, $-0.1<\rm{[Fe/H]}<0.1$, $0.1<\rm{[Fe/H]}<0.3$, and $0.3<\rm{[Fe/H]}<0.5$, respectively. The rightmost column shows the median results for different metallicity bins.}
    \label{feh_nuv.fig}
\end{figure*}

\begin{figure*}[!htbp]
\centering
    \includegraphics[width=0.98\textwidth]{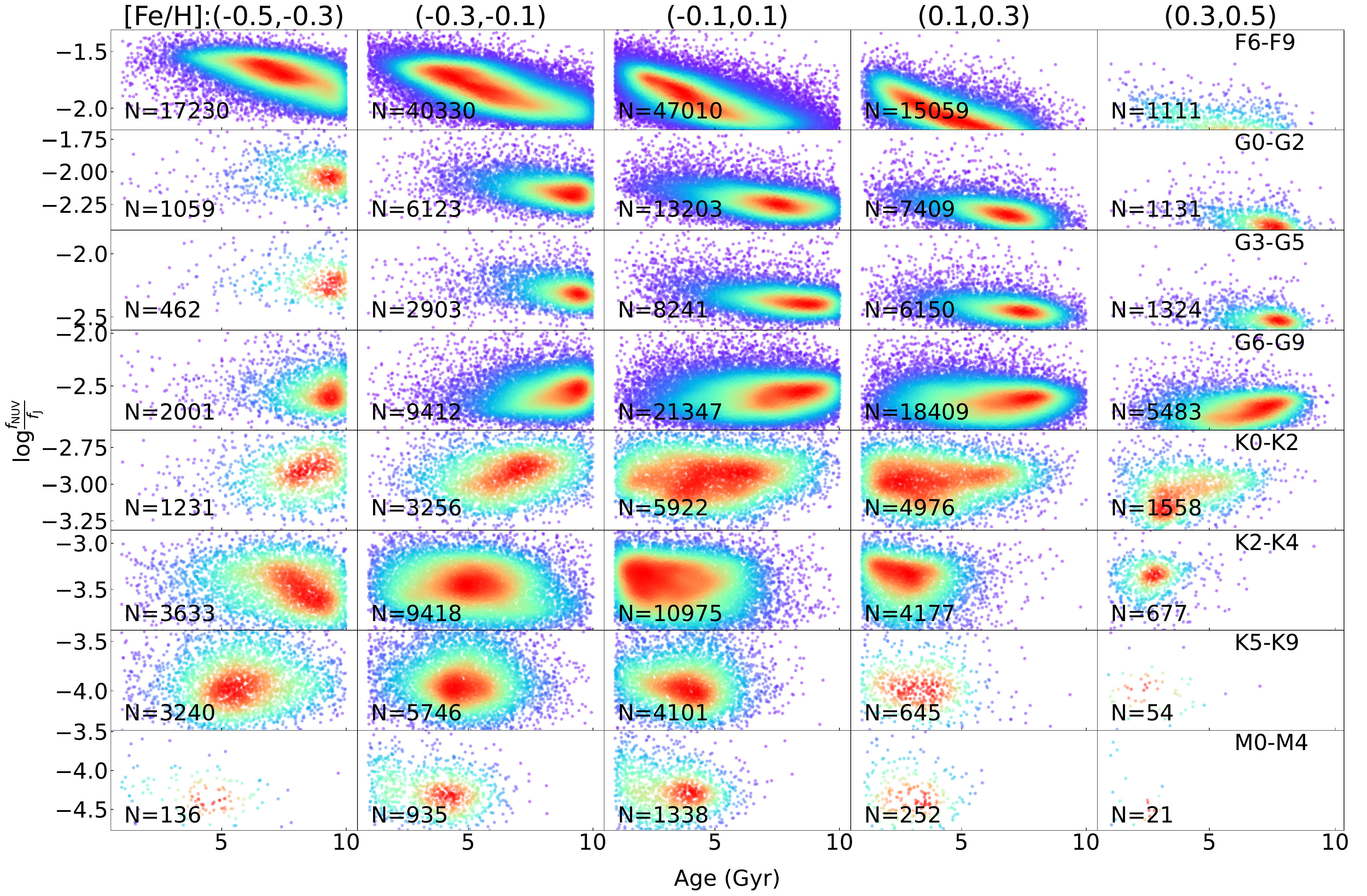}
    \caption{The relationship between NUV emission and stellar age for field stars ranging from spectral type A to M-dwarfs. The symbols are the same as shown in Figure \ref{feh_nuv.fig}. }
    \label{feh_nuv_field.fig}
\end{figure*}

The numerous studies have shown that metallicity can affect stellar activity measurement in blue band, especially for Ca II H\&K \citep[e.g.][]{2016A&A...594L...3L, 2024ApJS..271...19Y}. Here we explore the evolution of NUV emission with age for different spectral types of dwarfs, considering the impact of metallicity.
The metallicities for both the open cluster and field stars were adopted from the StarHorse catalog \citep{2019A&A...628A..94A}, which provided homogeneous estimates of stellar parameters by combining spectroscopic, photometric, and astrometric data.


We categorized the sample into five metallicity groups: $-0.5<\rm{[Fe/H]}<-0.3$, $-0.3<\rm{[Fe/H]}<-0.1$, $-0.1<\rm{[Fe/H]}<0.1$, $0.1<\rm{[Fe/H]}<0.3$, and $0.3<\rm{[Fe/H]}<0.5$. The relationships between NUV emission and stellar age for each spectral type across these metallicity categories are shown in Figures \ref{feh_nuv.fig} and \ref{feh_nuv_field.fig}. 
For each spectral type, metal-poor stars generally exhibit higher NUV emission than their solar-metallicity and metal-rich counterparts, consistent with previous findings based on Ca II H\&K activity studies \citep{2016A&A...594L...3L, 2024ApJS..271...19Y}. 
This trend is evident in both young stars in open clusters and older field stars, and can be attributed to the reduced line blanketing in metal-poor stars \citep{2013ApJ...766...69L,2014ApJ...780...61L}, which allows more NUV radiation to escape, resulting in enhanced NUV emission.

\subsection{Rotation period-age relation}
\label{uv_rot.sec}

\begin{figure}[!htbp]
    \centering
    \includegraphics[width=0.45\textwidth]{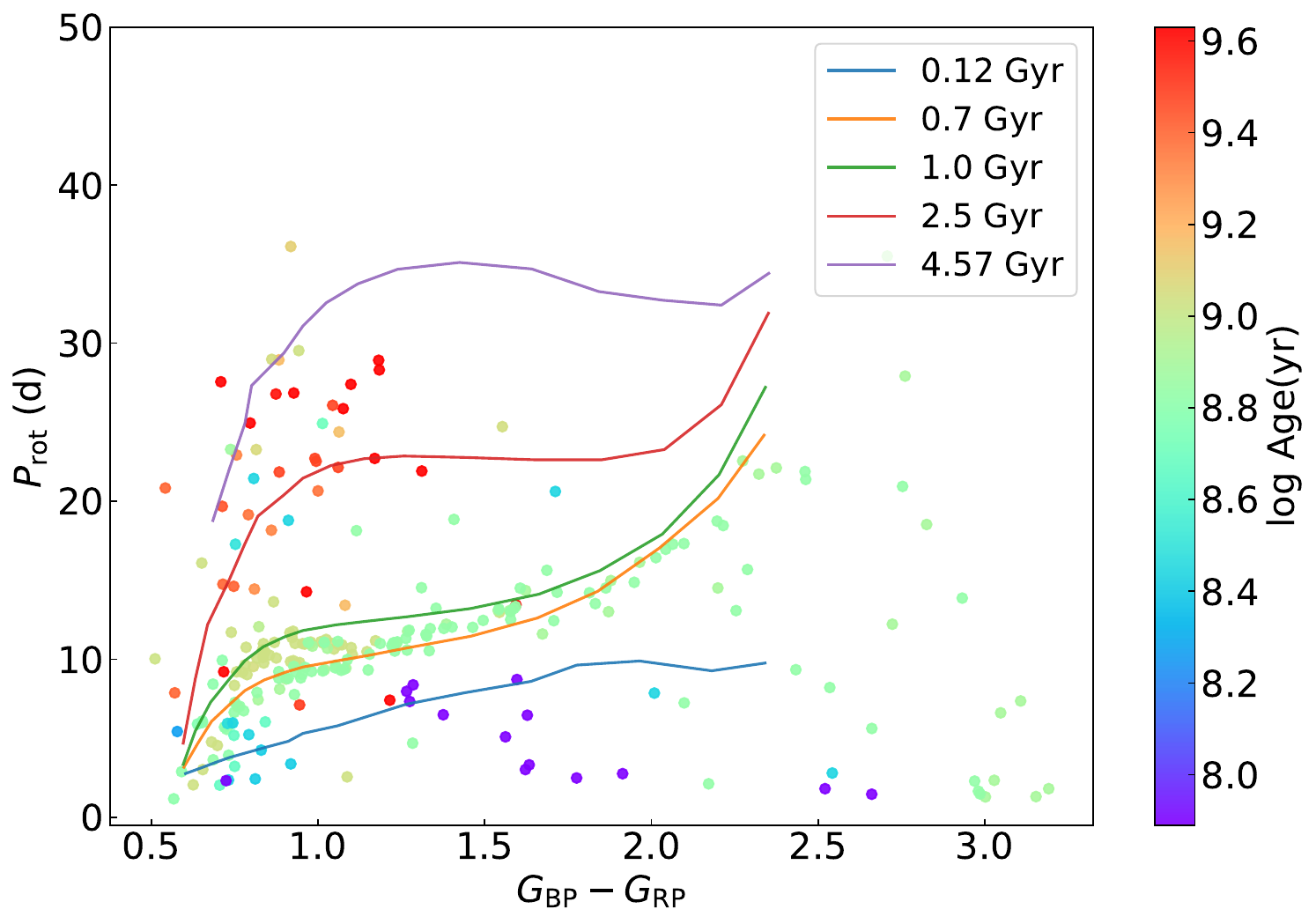}
    \caption{Rotation-Color Diagram. The color bar are the cluster age. The color lines are from the results by \cite{2020A&A...636A..76S}.}
    \label{rot.fig}
\end{figure}

During the main sequence phase, stars gradually lose angular momentum through magnetic braking, causing their rotation periods ($P_{\rm rot}$) to increase with ages. Over the past few decades, gyrochronology establishes a relationship between stellar rotation period, effective temperature, and age. This relationship allows for the estimation of stellar ages based on their rotational periods \citep{1972ApJ...171..565S, 2003ApJ...586..464B, 2007ApJ...669.1167B,2008ApJ...687.1264M}. Observations of open clusters such as the Pleiades, Hyades, and M67 have provided crucial constraints on rotational evolution models \citep{2016ApJ...822...47D, 2020A&A...636A..76S}. 

We cross-matched our open clusters sample with stars that have measured rotation periods from the Kepler survey \citep{2013MNRAS.432.1203M, 2014ApJS..211...24M} and the K2 mission \citep{2019ApJS..244...21S, 2020A&A...635A..43R}. We downloaded their light curves, folded them using the reported rotation periods, visually inspected them, and finally selected 250 stars with clear rotational modulation.

Figure \ref{rot.fig} presents the rotation–color diagram for all rotating open cluster stars in our sample. Several rotational isochrones from \cite{2020A&A...636A..76S} were also plotted, with ages at 0.12~Gyr, 0.7~Gyr, 1~Gyr, 2.5~Gyr, and 4.57~Gyr (approximately the age of the Sun). \cite{2020A&A...636A..76S} proposed a semi-empirical rotational evolution model based on physical mechanisms, which considers magnetic braking, angular momentum transport, and the coupling between the radiative core and convective envelope inside stars. Because \cite{2020A&A...636A..76S} provided the relationship between rotation and mass, we used the PARSEC model\footnote{\url{https://stev.oapd.inaf.it/cgi-bin/cmd}} to determine the corresponding $G_{\rm BP}-G_{\rm RP}$ for a given mass at a specific age, based on the solar metallicity. 

This relationship reflects how stellar rotation evolves over time. 
The rotation-color trend for stars around 1~Gyr aligns well with the relation reported by \cite{2020A&A...636A..76S}. The well-known stalling effect \citep{2020ApJ...904..140C} is evident for early K-type stars ($G_{\rm BP} - G_{\rm RP} \approx 1.2$) between $\sim 700$~Myr and $\sim 1$~Gyr.
On the other hand, older clusters like M67 (age $\sim 4.3$~Gyr) exhibit a more scattered rotation-color distribution that deviates from the relation of \cite{2020A&A...636A..76S}.
This deviation may result from the weakening of magnetic braking in older stars, leading to observable departures from the expected spin-down behavior \citep{2016Natur.529..181V, 2019ApJ...872..128V}. 
These findings suggest that although the rotation-age relation has been well established for stars younger than $\sim 2$~Gyr, further observational and theoretical work is needed to better understand stellar rotational evolution at older ages.

\section{Evolution of Stellar Habitable Zone}
\label{uhz.sec}

The boundaries of the habitable zone evolve with age due to variation in stellar radiation \citep{2024MNRAS.532..563S}. Traditionally, the habitable zone is defined based on the star’s luminosity and the distance at which a planet would experience surface temperature conducive to the presence of liquid water \citep{1993Icar..101..108K}, which is often called circumstellar habitable zone (CHZ).
Previous studies have refined the boundaries of the CHZ based on stellar properties and updated climate models \citep{2013ApJ...765..131K, 2014ApJ...787L..29K}, and further considering the planetary rotation rate \citep{2014ApJ...787L...2Y}. 

In recent years, the ultraviolet habitable zone (UHZ) has gained attention as a complement to the classical habitable zone, emphasizing the dual role of UV radiation in both supporting and hindering life \citep{2018SciA....4.3302R, 2023MNRAS.522.1411S}. 
On one hand, excessive levels of UV and X-ray radiation can strip planetary atmospheres and destroy essential biomolecules, particularly around active stars \citep{2007Icar..192..582B,2010A&A...511L...8S, 2023MNRAS.522.1411S, 2023ApJ...951...18M}. 
On the other hand, moderate UV radiation are thought to be crucial for prebiotic chemistry, driving the synthesis of key compounds like ribonucleic acid (RNA) in early planetary environments \citep{1977OrLi....8..259T, PMID:19444213, 2018SciA....4.3302R}.
Thus, UV radiation represents a double-edged sword for habitability—potentially harmful in excess, but possibly essential for the origin of life.

\subsection{Habitable zone calculation}
\label{hz_cal.sec}

We calculated the CHZ and UHZ evolution for different spectral types from F- to M-type dwarfs. We derived the CHZ for our sample across two age intervals: from log(Age/yr) = 6.5 to 9 (approximately 3~Myr to 1~Gyr) and from 1 to 10~Gyr (i.e., log(Age/yr) = 9 to 10), as presented in Section \ref{uv_type.sec} using the method described by \citet{2014ApJ...787L..29K}. 

The inner edge of the CHZ is determined based on the ``runaway greenhouse'' limit, where water-induced greenhouse effects become unsustainable. The outer edge is defined by the ``maximum greenhouse'' limit. The CHZ was calculated with the following equation \citep{2014ApJ...787L..29K}:
\begin{align}
d &= \left(\frac{L/L_{\odot}}{S_{\rm eff}}\right)^{0.5} \, \text{AU},
\end{align}
where $S_{\rm eff}$ represents the effective solar flux received by the Earth, defined as,
\begin{align}
S_{\rm eff} &= S_{\rm eff,0} + aT + bT^2 + cT^3 + dT^4.
\end{align}
In these calculations, we assumed the planet’s mass to be equal to that of the Earth.
$T$ is defined as $T_{\rm eff}-5870 \rm~K$.

(1) Figure \ref{uhz.fig} (left panel) displays the distribution of CHZ for all of our sample stars (marked by short gray vertical lines).

(2) We also calculated the theoretical CHZ for different types of dwarfs with stellar models. 
First, for each spectral type, we selected three different masses\footnote{\url{https://www.pas.rochester.edu/~emamajek/EEM_dwarf_UBVIJHK_colors_Teff.txt}}---early (e.g., F0), median (e.g., F4), and late (e.g., F9) stars---as representatives for the calculation.
Second, for each mass, we picked up its effective temperature and luminosity at each evolution phase (before the red giant branch), using the MIST EEP model.
These parameters were then used for the calculations of CHZ.
Moreover, we also calculated the theoretical CHZ taking into account the effects of planetary rotation rates following \citet{2014ApJ...787L...2Y}, which utilized the Community Atmosphere Model version 3.1 (CAM3) with a sophisticated cloud scheme. These simulations highlight that the inner edge of the CHZ is significantly influenced by the planet's rotation rate. Slowly rotating planets, previously deemed uninhabitable, could reside within the CHZ under certain conditions.

We calculated the boundaries of the UHZ using GALEX NUV luminosities, as defined by \cite{2023MNRAS.522.1411S}. Here we use the luminosity obtained by fitting the sample after considering the distance completeness as the luminosity to calculate the UHZ. The outer boundary of the UHZ corresponds to a NUV flux of $f_{\rm NUV} \geq 45~\text{erg cm}^{-2}~\text{s}^{-1}$, which represents the minimum NUV flux required for abiogenesis \citep{2018SciA....4.3302R}. The inner boundary is set at $f_{\rm NUV} \leq 1.04 \times 10^4~\text{erg cm}^{-2}~\text{s}^{-1}$, approximately twice the NUV radiation received by the Earth during the Archean eon ($\sim 3.8$~Gyr). 
We also accounted for the impact of varying atmospheric absorption on the NUV flux reaching a planet's surface. Flux values were scaled to $10\%$, $50\%$, and 100\% of the top-of-atmosphere flux, corresponding to atmospheric UV transparency values ($f$) of 0.1, 0.5, and 1.0, respectively.

\subsection{Evolution of CHZ and UHZ}
\label{hz_evo.sec}

\begin{figure*}[!htbp]
\centering
    \includegraphics[width=0.98\textwidth]{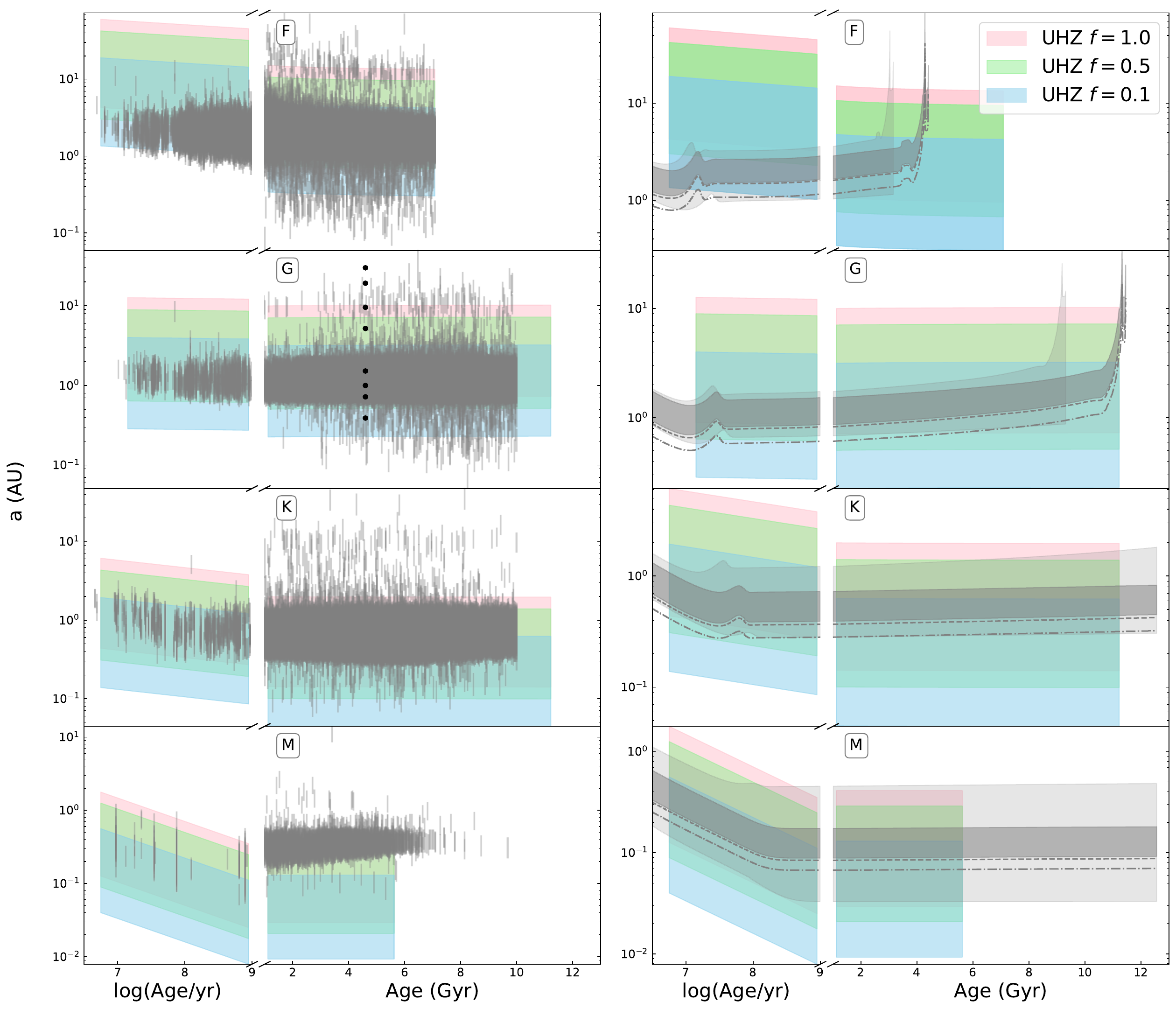}
    \caption{Left Panel: The CHZ and UHZ evolution of different spectral types from F- to M-dwarfs. The gray vertical lines are the CHZ range of each sources. The pink, green, and blue shaded areas represent the UHZ for different NUV transmission rates of planetary atmospheres. The black dots represent the planets in our Solar System.
    Right Panel: The CHZ and UHZ evolution of different spectral types from F- to M-dwarfs. The gray shaded areas are the CHZ results obtained using the mass calculations of med-type stars from MIST model, and the lighter gray shaded areas are the wider CHZ ranges obtained using the mass calculations of early- and late-type stars. The gray dashed line and gray dot line are the inner boundaries of CHZ, which take into account the rotation rates of different stars. The other symbols are the same as the left panel. }
    \label{uhz.fig}
\end{figure*}

The CHZ and UHZ evolution are shown in Figure \ref{uhz.fig}. The CHZ evolution from the models of F, G, and K stars shows a characteristic bump at log(age) $\sim$ 7, 7.5, and 8 (Figure \ref{uhz.fig}, right panel). This bump corresponds to the pre-main sequence phase, during which stars contract and heat rapidly, causing sharp changes in luminosity and temperature \citep{2015A&A...577A..42B}. As they settle onto the main sequence, these properties stabilize, and the CHZ shifts inward accordingly. After the main-sequence phase, F and G stars evolve quickly into sub-giants or giants, driving a rapid outward shift in their CHZ. In contrast, M dwarfs evolve more slowly and smoothly, with the CHZ gradually moving inward during their pre-main sequence phase.

By comparing the CHZ and UHZ, we found that, under the most optimistic scenario, planets within the habitable zone could be found around stars of all spectral types from F to M. However, for F dwarfs, CHZ and UHZ only coincide when the transmission is 0.1. This implies that strong atmospheric attenuation is required to maintain biologically safe NUV levels on the planetary surface. G and K stars show very stable UHZ and CHZ during the main sequence, which are more promising candidates for finding habitable planets around them. Without considering planetary migration and other dynamical factors, the stable habitable zones of G- and K-type stars may offer more favorable environments for NUV-triggered origin of life scenarios, similar to that of the early Earth.

In our sample, we include only early-type M stars.
For M stars in the pre-main-sequence phase, both the observed (Figure \ref{uhz.fig}, left panel) and theoretical (Figure \ref{uhz.fig}, right panel) distributions of the CHZ match well with the UHZ.
However, in the main-sequence phase, the observed CHZ deviates from the UHZ and only aligns with it under conditions of high UV transparency ($f = 1.0$). This suggests that planets around M stars may only remain habitable during their pre-main-sequence phase.


This finding is consistent with the conclusions of \cite{2024MNRAS.533L..76S}, which focuses on individual stars hosting planets. In contrast, our results provide general evolutionary trends for the habitable zones of M-type stars. It is important to note that the habitable zone of specific stars may vary depending on their individual stellar properties and planetary system dynamics.

We remind that the effect of stellar activity variability on the habitable zone is not considered in this study. Our previous work on M dwarfs \citep{2024ApJ...966...69L} and exoplanet host stars \citep{2025ApJS..276...29L} has shown that UV variability has a negligible impact on the habitable zone in a long timescale. Even large flares, provided they are not highly frequent, have durations of only a few to tens of minutes and therefore do not significantly alter the habitable zone. However, our analysis only considered the total luminosity from electromagnetic radiation. Strong radiation, proton events, and coronal mass ejections associated with flares may significantly affect planetary atmospheres and, in turn, habitability, but this remains an open question \citep{2010AsBio..10..751S, 2019AsBio..19...64T}.

\section{Summary}
\label{summary.sec}

To investigate the evolution of the NUV emission across different stellar types, we compiled a catalog of open cluster members and field stars with age estimates. The ages of open clusters were obtained from isochrone fitting in previous studies, while the ages of field stars were derived using a data-driven approach based on LAMOST spectra. In total, we selected 386,500 main-sequence stars with NUV magnitude measurements, including 15,938 A- to M-type dwarfs from open clusters (aging from 3~Myr to 1~Gyr) and 370,562 F- to M-type dwarfs from the field (aging from 1 to 10~Gyr).

Our results show that NUV emission evolves differently across spectral types. For A- and early F-type stars, the emission is dominated by thermal (photospheric) radiation, and its gradual decline is primarily driven by changing in stellar radius over time. In contrast, the NUV emission of late F- to M-type stars is strongly influenced by magnetic field. 
Late F- and G-type stars show a decline in NUV emission after $\sim$1~Gyr, consistent with the spin-down due to magnetic braking. Meanwhile, K- and M-type stars exhibit a rapid drop in NUV emission between from 7~Myr to 1~Gyr, reflecting the transition from the YSO phase to the main-sequence phase.

We confirmed that the rotation–color–age relationship of our sample stars agrees well with the predictions by the core-envelope decoupling model \citep{2020A&A...636A..76S} for stars younger than 2~Gyr. However, a deeper understanding of stellar rotational evolution requires more rotation measurements for older stars (e.g, those older than 4~Gyr).

We also tracked the evolution of the UHZ and compared it with the CHZ across different spectral types. Our results show that for F-type dwarfs, the CHZ and UHZ coincide only when the atmospheric UV transparency is very low. G- and K-type stars provide the most stable overlap between the two zones throughout their long-term evolution. In contrast, for M-type stars, the CHZ and UHZ match well during the pre-main-sequence phase, but start to separate during the main sequence, potentially resulting in an uninhabitable environment.

\section*{Acknowledgements}

We thank the anonymous referee for helpful comments and suggestions that have significantly improved the paper. 
GALEX data presented in this paper were obtained from the Mikulski Archive for Space Telescopes (MAST).
The Guoshoujing Telescope (the Large Sky Area Multi-Object Fiber Spectroscopic Telescope LAMOST) is a National Major Scientific Project built by the Chinese Academy of Sciences. Funding for the project has been provided by the National Development and Reform Commission. LAMOST is operated and managed by the National Astronomical Observatories, Chinese Academy of Sciences.
This work presents results from the European Space Agency (ESA) space mission {\it Gaia}. {\it Gaia} data are being processed by the {\it Gaia} Data Processing and Analysis Consortium (DPAC). Funding for the DPAC is provided by national institutions, in particular the institutions participating in the {\it Gaia} MultiLateral Agreement (MLA). We acknowledge use of the VizieR access tool, operated at CDS, Strasbourg, France, and of Astropy, a community-developed core Python package for Astronomy (Astropy Collaboration, 2013). 
This work was supported by National Natural Science Foundation of China (NSFC) under grant Nos. 12588202/12273057/11833002/12090042, the National Key Research and Development Program of China (NKRDPC) under grant number 2023YFA1607901, the Strategic Priority Program of the Chinese Academy of Sciences under grant number XDB1160302, and science research grants from the China Manned Space Project. J.F.L acknowledges the support from the New Cornerstone Science Foundation through the New Cornerstone Investigator Program and the XPLORER PRIZE.

\bibliographystyle{yahapj}
\bibliography{bibtex.bib}{}

\begin{appendix}

\section{Comparison of different open clusters catalogs}
\label{comOC.sec}
\renewcommand{\thefigure}{A\arabic{figure}}
\setcounter{figure}{0}

\begin{figure*}[!htbp]
    \centering
    \includegraphics[width=0.98\textwidth]{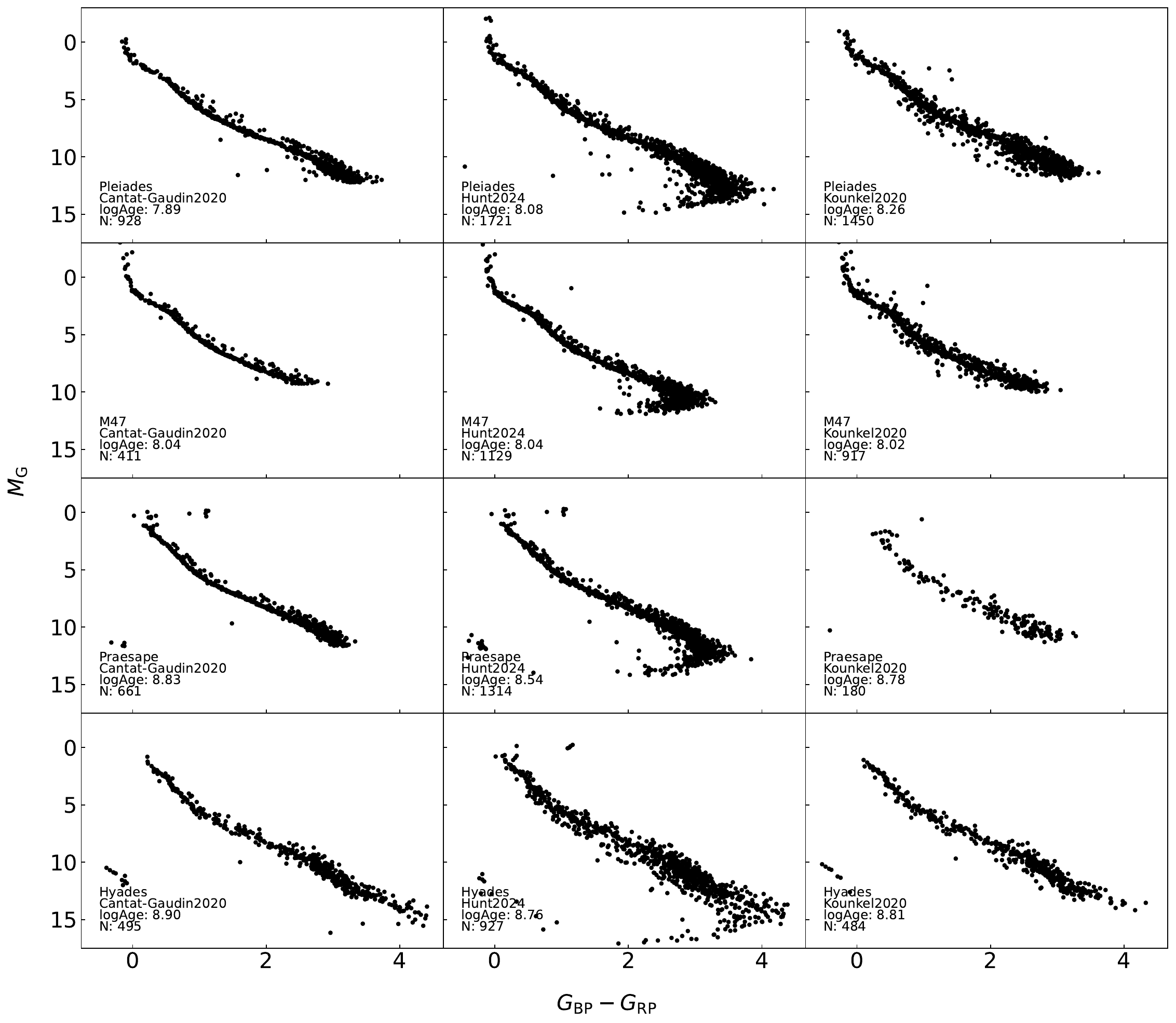}
    \caption{Four examples of cluster CMD from the different datasets. From top to bottom, the clusters are Pleiades, M47, Praesepe, and Hyades. From left to right, the data are sourced from \cite{2020A&A...640A...1C}, \cite{2024A&A...686A..42H} and \cite{2020AJ....160..279K}.}
    \label{hr_4_clusters_compare.fig}
\end{figure*}

\begin{figure*}[!htbp]
    \centering
    \includegraphics[width=0.98\textwidth]{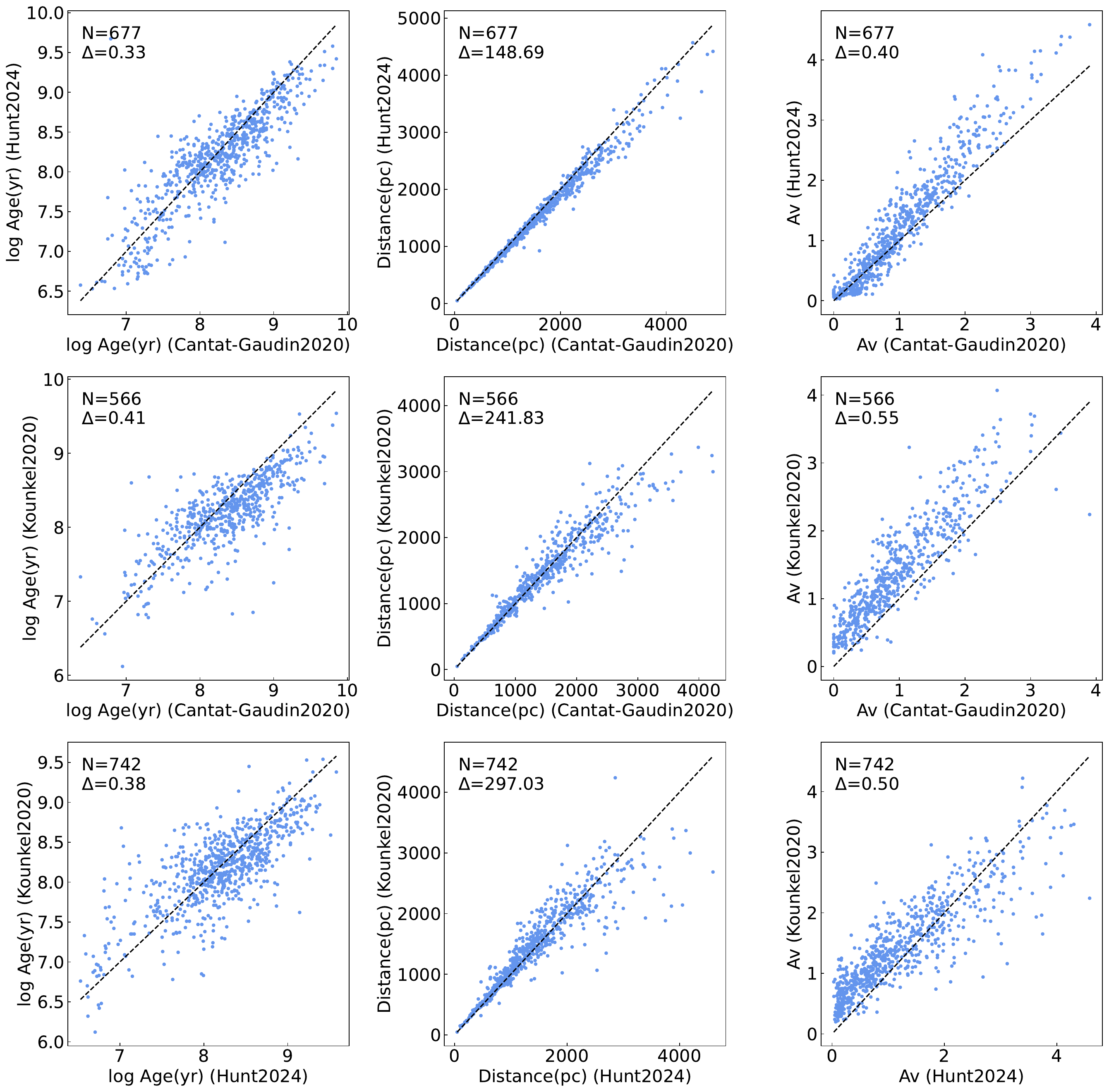}
    \caption{Comparison of parameters given between different datasets. From left to right are age, distance and extinction. The dashed black line is y=x.}
    \label{oc_pars_compare.fig}
\end{figure*}

Figures \ref{hr_4_clusters_compare.fig} and \ref{oc_pars_compare.fig} show comparisons of CMD and stellar parameters across different open clusters catalogs.

\clearpage

\section{The evolution of all main sequences}
\label{all_evo.sec}
\renewcommand{\thefigure}{B\arabic{figure}}
\setcounter{figure}{0}

\begin{figure}[!htbp]
    \centering
    \includegraphics[width=0.98\textwidth]{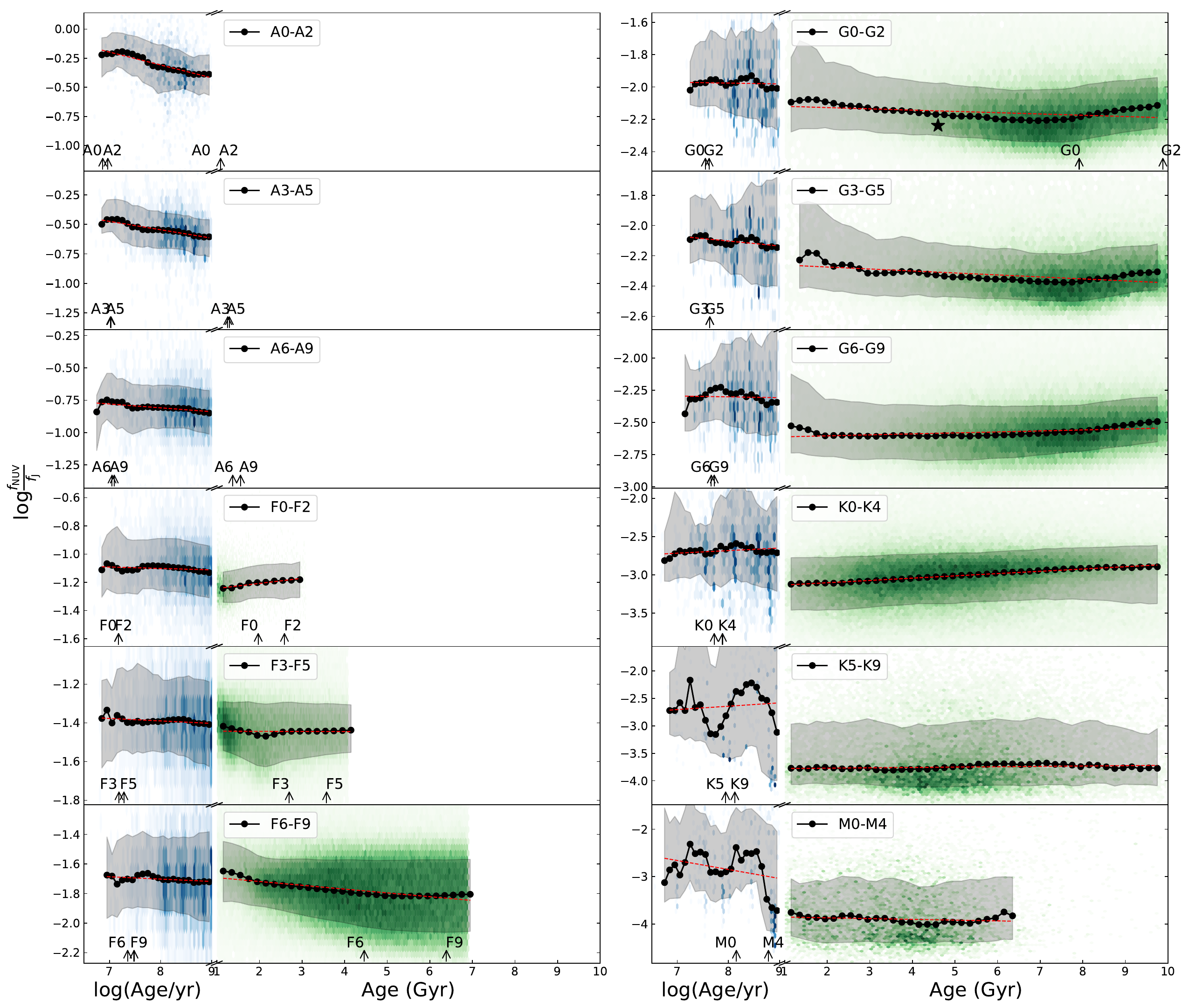}
    \caption{The relation between the log$f_{\rm NUV}/f_{\rm J}$ and stellar age of all main sequences. The meanings of different symbols are consistent with those in Figure \ref{distribution_fnuv_oc_field.fig}.}
    \label{distribution_fnuv_all.fig}
\end{figure}

\begin{figure}[!htbp]
    \centering
    \includegraphics[width=0.98\textwidth]{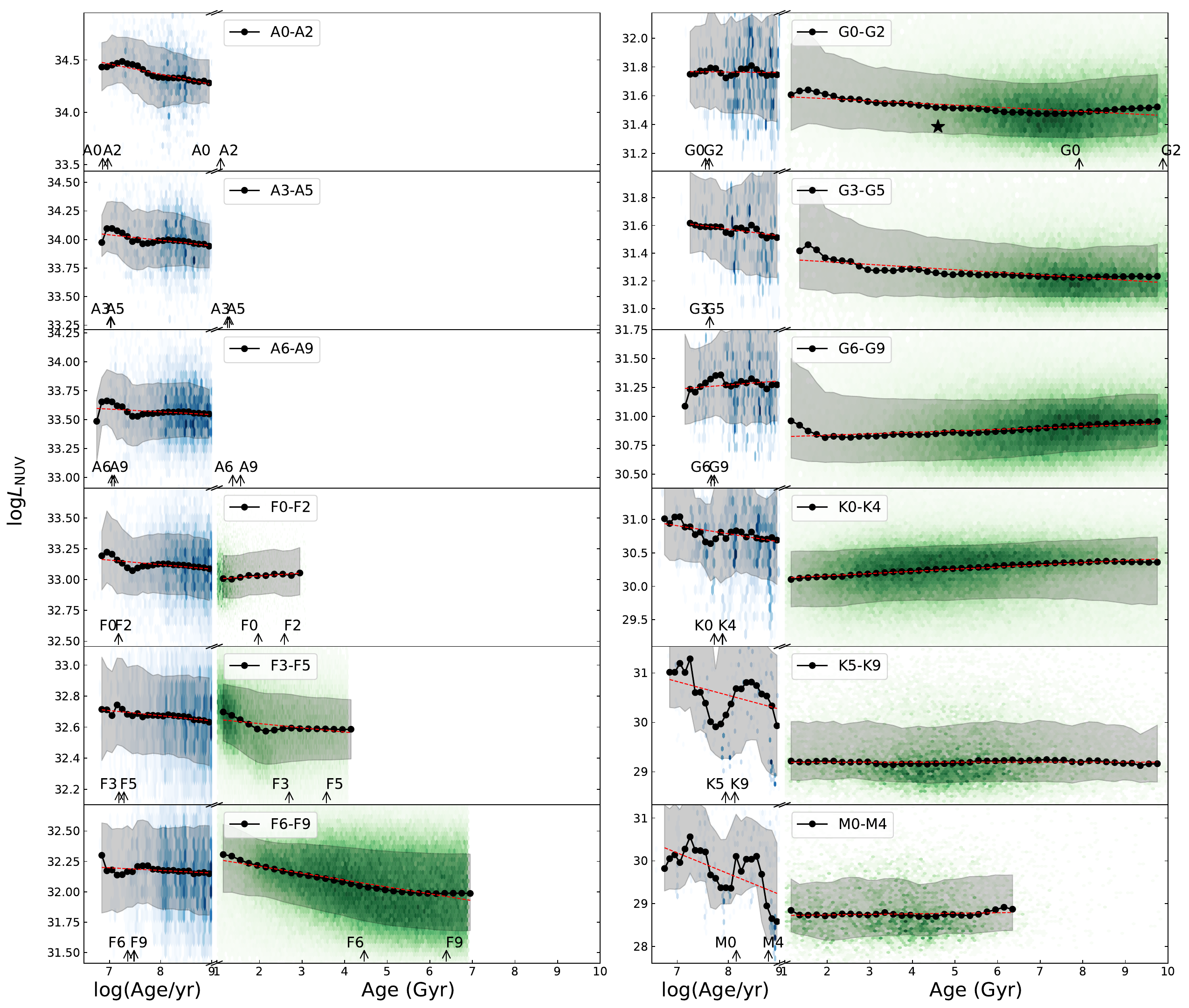}
    \caption{The relation between the log$L_{\rm NUV}$ and stellar age of all main sequences. The meanings of different symbols are consistent with those in Figure \ref{distribution_fnuv_oc_field.fig}.}
    \label{distribution_lnuv_all.fig}
\end{figure}

\begin{figure}[!htbp]
    \centering
    \includegraphics[width=0.98\textwidth]{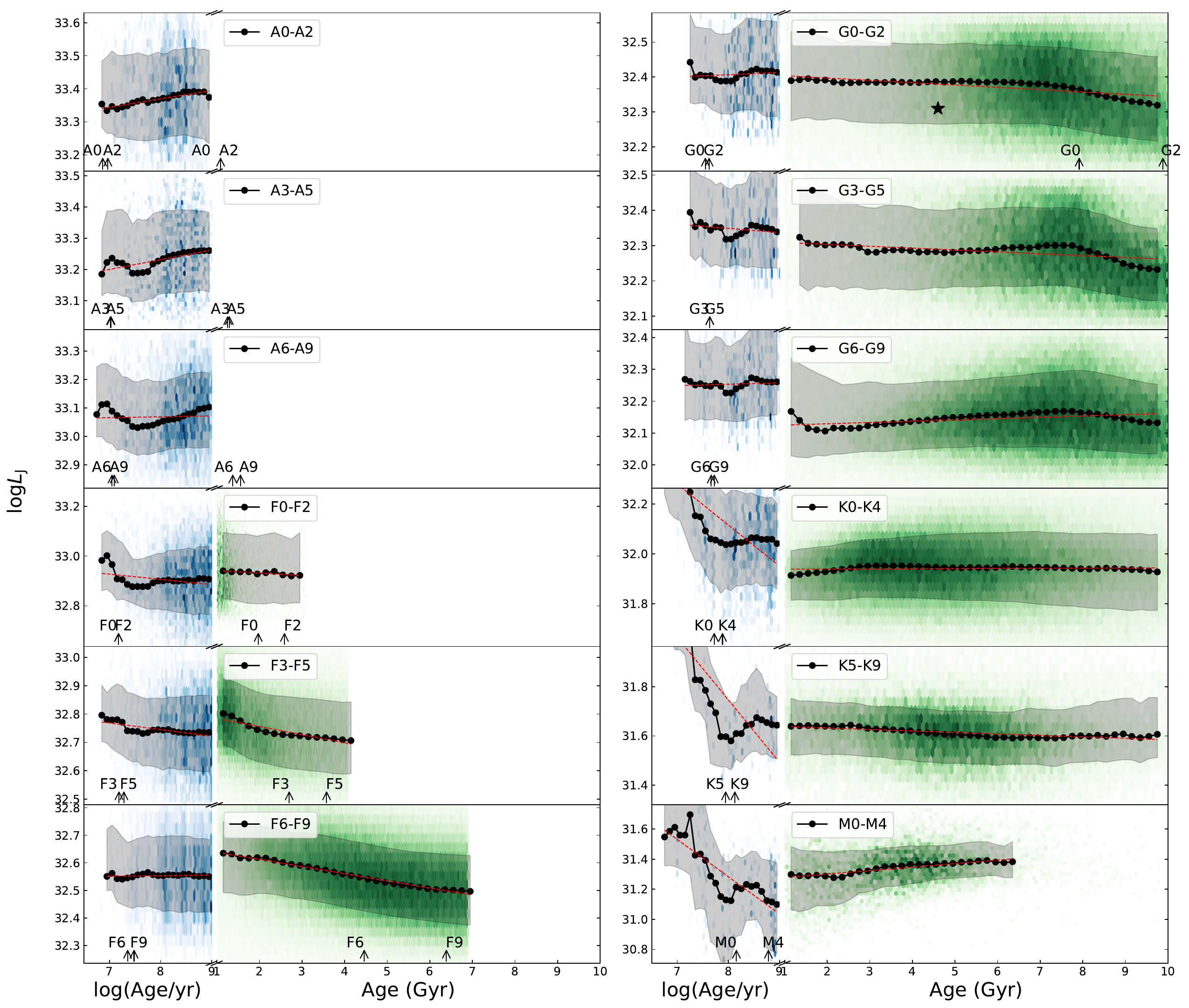}
    \caption{The relation between the log$L_{\rm J}$ and stellar age of all main sequences. The meanings of different symbols are consistent with those in Figure \ref{distribution_fnuv_oc_field.fig}.}
    \label{distribution_lj_all.fig}
\end{figure}

Figure \ref{distribution_fnuv_all.fig}, \ref{distribution_lnuv_all.fig}, and \ref{distribution_lj_all.fig} show the evolution of NUV and J-band emission for the total sample without considering distance completeness.

\end{appendix}

\end{document}